\documentclass[twocolumn]{aastex63}
 
\usepackage[T1]{fontenc}
\usepackage{ae,aecompl}
\usepackage{graphicx}
\usepackage{pgfplots}
\usepackage{mathtools}
\usepackage{epstopdf}
\usepackage{graphicx}	
\usepackage{amsmath}	
\usepackage{amssymb}
\usepackage{subfigure}
\usepackage{svg}
\usepackage{xcolor}
\usepackage{afterpage}
\usepackage{booktabs}
\usepackage{float}
\usepackage{upgreek}

\received{January 13, 2022}
\revised{April 13, 2022}
\accepted{April 23, 2022}
\submitjournal{ApJ}

\shorttitle{Orbital parameters of PSR~J1603$-$7202}
\shortauthors{Walker et al.}

\begin{document}
\title{Orbital dynamics and extreme scattering event properties from long-term scintillation observations of PSR~J1603$-$7202}

\correspondingauthor{Kris Walker}
%\suppressAffiliations
\email{kris.walker@icrar.org}
\author{Kris Walker}
\affiliation{Centre for Astrophysics and Supercomputing, Swinburne University of Technology, P.O. Box 218, Hawthorn, Victoria 3122, Australia}
\affiliation{Monash Centre for Astrophysics (MoCA), School of Physics and Astronomy, Monash University, Victoria 3800, Australia}
\affiliation{Australian Research Council Centre of Excellence for Gravitational Wave Discovery (OzGrav)}
\author{Daniel J. Reardon}
\affiliation{Centre for Astrophysics and Supercomputing, Swinburne University of Technology, P.O. Box 218, Hawthorn, Victoria 3122, Australia}
\affiliation{Australian Research Council Centre of Excellence for Gravitational Wave Discovery (OzGrav)}

\author{Eric Thrane}
\affiliation{Monash Centre for Astrophysics (MoCA), School of Physics and Astronomy, Monash University, Victoria 3800, Australia}
\affiliation{Australian Research Council Centre of Excellence for Gravitational Wave Discovery (OzGrav)}

\author{Rory Smith}
\affiliation{Monash Centre for Astrophysics (MoCA), School of Physics and Astronomy, Monash University, Victoria 3800, Australia}
\affiliation{Australian Research Council Centre of Excellence for Gravitational Wave Discovery (OzGrav)}

\begin{abstract}
We model long-term variations in the scintillation of binary pulsar PSR J1603$-$7202, observed by the 64\,m Parkes radio telescope (Murriyang) between 2004 and 2016. We find that the time variation in the scintillation arc curvature is well-modelled by scattering from an anisotropic thin screen of plasma between the Earth and the pulsar. Using our scintillation model, we measure the inclination angle and longitude of ascending node of the orbit, yielding a significant improvement over the constraints from pulsar timing. From our measurement of the inclination angle, we place a lower bound on the mass of J1603$-$7202's companion of $\gtrsim 0.5\,\text{M}_\odot$ assuming a pulsar mass of $\gtrsim1.2\,\text{M}_\odot$. We find that the scintillation arcs are most pronounced when the electron column density along the line of sight is increased, and that arcs are present during a known extreme scattering event. We measure the distance to the interstellar plasma and its velocity, and we discuss some structures seen in individual scintillation arcs within the context of our model.
\end{abstract}

\section{Introduction} 
\label{sec:intro}

%Scattering and scintillation. Papers to start: Rickett 1990, Cordes and Rickett 1998, Narayan 1992
%
%Scintillation arcs. Papers to start: Stinebring 2001, Walker 2004, Cordes 2006
%
%Annual variations in long term scintillation models. Papers to start: Rickett 2014, Reardon 2019, Reardon 2020
%
%Millisecond pulsar timing, and DM variations in J1603. Papers to start: Manchester 2013, Keith 2013, Coles 2015, Reardon 2016, Kerr 2020

\begin{figure*}
    \centering
    \includegraphics[width=0.9\textwidth]{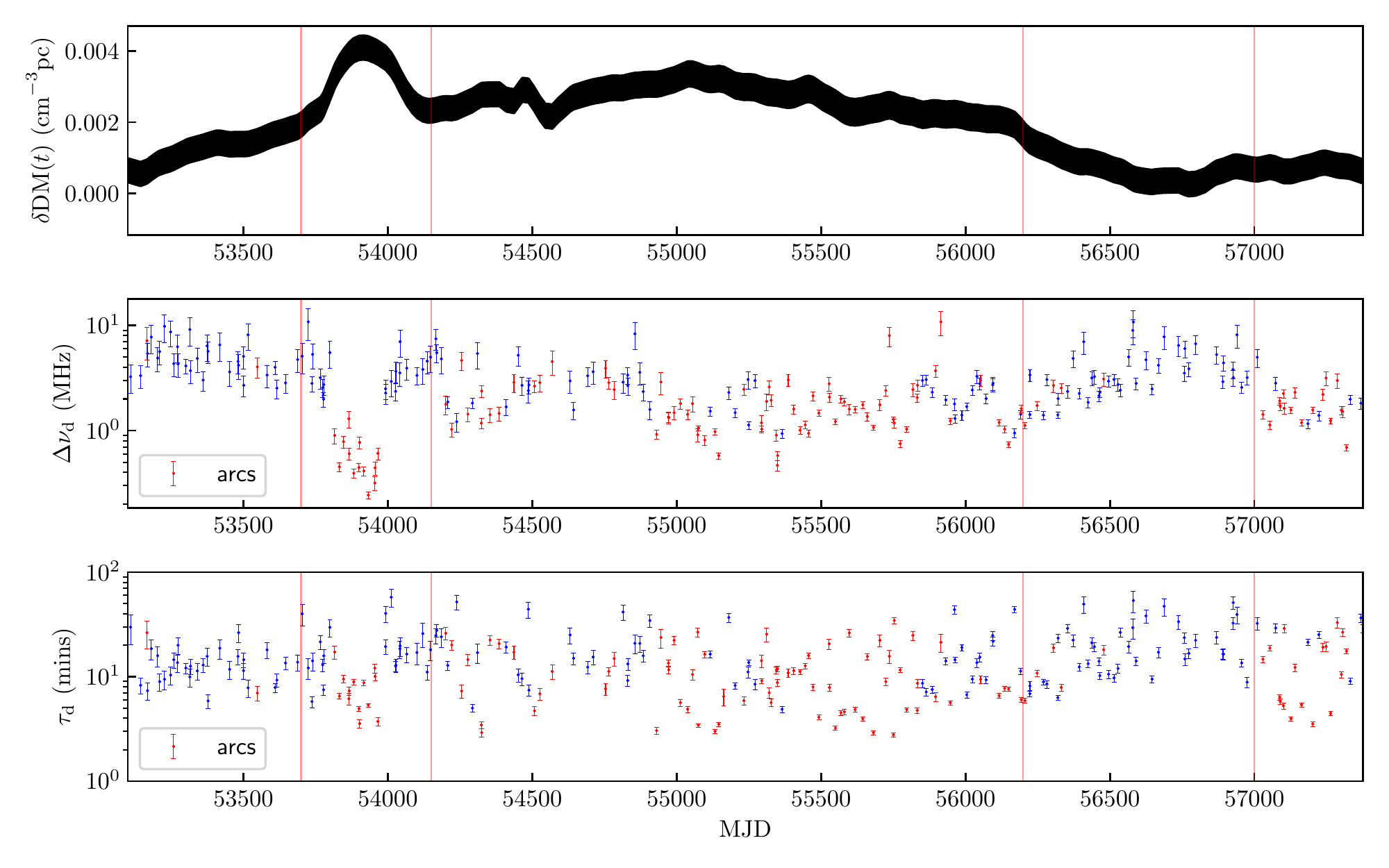}
    \caption{Time variations in the dispersion measure (top), decorrelation bandwidth $\Delta\nu_\text{d}$ (middle), and scintillation timescale $\tau_\text{d}$ (bottom) for J1603$-$7202. Both the decorrelation bandwidth and scintillation timescale are inversely correlated with the DM variations and the ESE at around MJD 53900 is visible in all plots. The red data points correspond to observations that exhibit prominent scintillation arcs, and their distribution reveals a correlation with the dispersion measure. The vertical red lines indicate the regions of notable DM variation: the first two encompass the ESE, which is followed by $\sim 2,000$ days of enhanced DM before a steady decrease (and a corresponding increase in $\Delta\nu_\text{d}$).}
    \label{fig:dm_variations}
\end{figure*}

Observations of the interstellar medium (ISM) over a wide range of spatial scales reveal a turbulent cascade over a remarkable twelve orders of magnitude in wavenumber \citep[for a review, see][]{Elmegreen+04}. At small scales, large frequency-dependent brightness variations in pulsar signals suggest scattering by overdense AU-scale concentrations of turbulent plasma, known as extreme scattering events (ESEs) \citep{Fiedler+94}. The formation mechanisms and energy sources required to sustain these overdensities remain unclear, though several hypotheses have been proposed, such as local density enhancements \citep{Walker+98}, highly asymmetric structures like current sheets \citep{Pen+12}, and circumstellar plasma streams \citep{Walker+17}. One pulsar that underwent such a scattering event is PSR~J1603$-$7202, a binary millisecond pulsar with an orbital period of 6.31 days. The 2006 extreme scattering event was marked by a spike in the dispersion measure (DM) on the order of $10^{-3}\,\text{cm}^{-3}\text{pc}$, which was followed by a period of enhanced DM that lasted $\sim$ $2,000$ days \citep{Coles+15}. Dispersion is observed as a frequency-dependent group delay to pulse arrival times, and is related to the integrated column density of electrons along the line of sight to the pulsar at some distance $d$ by
\begin{align}
    \mathrm{DM} = \int_0^d n_\text{e}(s)\,\mathrm{d}s.
\end{align}
The 2006 increase in DM is therefore suggestive of a dense cell of plasma passing between the Earth and the pulsar. The variations in DM with time are shown in the top panel of Figure \ref{fig:dm_variations}, with the ESE spike and following period of enhanced DM marked by the regions between the red lines.

\begin{figure*}
    \centering
    \includegraphics[scale=0.7]{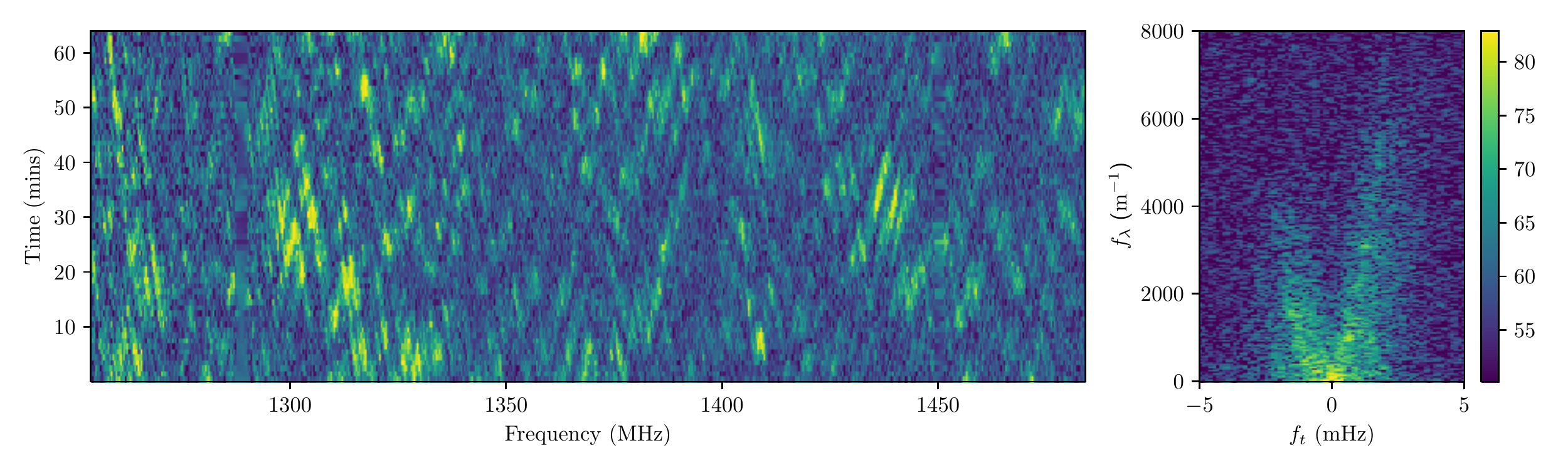}
    \caption{Example of a dynamic spectrum (left) and its corresponding secondary spectrum, featuring a prominent scintillation arc (right).}
    \label{fig:scint_obs}
\end{figure*}

To better understand the nature of these overdensities we require accurate measurements of the ISM structure responsible for the scattering. One promising avenue for probing the ISM at the small scales required is pulsar scintillation. When the pulsar radiation is scattered by density inhomogeneities in the ionized ISM, the diffracted wavefronts interfere and form a pattern of intensity variations in frequency and time at the observer \citep{Rickett69}, which can be displayed as a ``dynamic spectrum" (see Figure \ref{fig:scint_obs}). The characteristic scales of these variations are denoted as the decorrelation bandwidth, $\Delta\nu_\text{d}$, and scintillation timescale, $\tau_\text{d}$ and they arise as a result of the combined motions of the pulsar, Earth, and scattering plasma \citep{Lyne+84}, as well as changes in the structure of the plasma \citep{Cordes+98}. Periodic patterns in the dynamic spectrum, thought to be caused by interference between separate scattered images, map to discrete features in the associated ``secondary spectrum.'' Calculated by taking the squared magnitude of the 2D Fourier transform (the power spectrum) of the dynamic spectrum, the secondary spectrum describes the flux as a function of differential time delay and differential Doppler shift between pairs of interfering waves. Since \citet{Stinebring+01}, it has been known that secondary spectra sometimes exhibit power distributed in striking parabolic ``scintillation arcs" that result from ``criss-cross" patterns in the dynamic spectra. Theoretical treatments suggest that the distribution of power in these secondary spectra are directly related to on-sky angular brightness distributions \citep{Walker+04,Cordes+06}. It may therefore be possible to use these spectra to image the physical scattering structure. For PSR~J1603$-$7202, we observe many such scintillation arcs, with their appearance correlated with the dispersion measure variations. This can be seen clearly in a time series of the decorrelation bandwidth, which is inversely correlated with the DM. Highlighting those observations that feature prominent scintillation arcs reveals that enhancements in the DM are accompanied by the appearance of such arcs (see middle panel of Figure \ref{fig:dm_variations}). A similar but weaker correlation is also observed in the scintillation timescale (bottom panel of Figure \ref{fig:dm_variations}). This identifies the extreme scattering with the same scintillating plasma responsible for the scintillation arcs.

Unfortunately, relating scintillation observations to on-sky scattered images is complicated by the fact that the transformation requires knowledge of the pulsar's velocity and the distance and velocity of the scattering structure \citep{Walker+04}. Thankfully, scintillation arcs offer a way to determine these quantities through modelling variations in their curvatures. The curvature of a scintillation arc is related to the distances to the pulsar and scattering plasma, and the combined velocities of the pulsar, scattering plasma, and Earth. For pulsars undergoing binary orbital motion, variations in the curvature can be used to determine the orbital parameters. Long-term analyses of orbital and annual variations using scintillation arcs have only been performed twice: the results from \citet{Stinebring+05} for PSR J0737$-$3039 were consistent with but inferior to those from scintillation timescale variations, while \citet{Reardon+20} was able to precisely determine the orbit of PSR J0437$-$4715, providing an even better value for the longitude of ascending node than pulsar timing. Arc curvature variations are a robust measure of pulsar and ISM motion in that they are a purely geometric quantity, independent of the strength of scintillation \citep{Cordes+06} (see Section \ref{sec:curvatures} for a definition). This is in contrast to variations in the decorrelation bandwidth and scintillation timescale \citep{Rickett+90}, which have been used to model the velocities on several occasions \citep[e.g.][]{Lyne+84,Ord+02a,Rickett+14,Reardon+19}.

Regular observations of PSR~J1603$-$7202 have been carried out using the Parkes telescope since 2004 as part of the PPTA project \citep{Manchester+13}, providing us with an extensive archive of dynamic spectra that feature scintillation arcs in their power spectra spanning more than 10 years. In this paper we show that the arc curvature variations of J1603$-$7202 are well-modelled by an anisotropic thin screen scattering geometry, allowing us to determine the pulsar's orbital parameters and properties of the scattering screen. Section \ref{sec:observations} describes the observations and basic processing performed to obtain our dynamic and secondary spectra. Section \ref{sec:model} follows with a description of our models for the arc curvature and velocity, as well as the transformation from the secondary spectra to curvature probability distributions. Section \ref{sec:method} details our approach to Bayesian modelling of the data, the results of which are described in Section \ref{sec:results}. In Section \ref{sec:discussion} we discuss the agreement with pulsar timing, interpret physically the results for the scattering screen parameters, and place a lower bound on the companion mass, before discussing the interpretation of a selection of interesting secondary spectra.

\section{Observations}
\label{sec:observations}

PSR~J1603$-$7202 is a target of the Parkes Pulsar Timing Array (PPTA) project \citep{Manchester+13}, which performs observations of the pulsar on average every two weeks, using the Parkes 64 m radio telescope (Murriyang). For this work, we use observations from the PPTA data release 2 \citep{Kerr+20}, spanning more than a decade: from June 2005 to December 2015 (MJD 53548 to 57376). The observations were taken in three separate observing bands: 40/50-cm (at centre frequencies $\nu_\mathrm{c}\sim 732$ MHz and $\nu_\mathrm{c}\sim 685$ MHz respectively), 20-cm ($\nu_\mathrm{c}\sim 1400$ MHz), and 10-cm ($\nu_\mathrm{c}\sim 3100$ MHz). We only use observations in the 20-cm band since it is the only band in which scintillation arcs are resolved. Details of the observing systems are described in \citet{Manchester+13} and raw data processing in \citet{Kerr+20}.

\subsection{Dynamic and secondary spectra}
The dynamic spectra, $S(t,\nu)$, are computed as part of the data processing pipeline developed for PPTA data release 2, which uses the \textsc{psrchive} package \citep{Hotan+04}. The dynamic spectra are further processed using the \textsc{scintools}\footnote{\url{https://github.com/danielreardon/scintools}} Python package \citep{Reardon+20}. The zero-valued band edges of the dynamic spectrum are trimmed and additional artefacts in the dynamic spectrum at a $\geq 7\sigma$ deviation are zeroed and refilled using linear interpolation. This removes any potential radio-frequency interference and reduces artefacts in the secondary spectra (particularly along the axes) while leaving the curvatures of the scintillation arcs unaffected.

Before generating the secondary spectra, the dynamic spectra are resampled into equal steps of wavelength rather than frequency, $S(t,\nu)\rightarrow S(t,\lambda)$ \citep[as in][]{Reardon+21}. This avoids the frequency-dependence of the arc curvature and so allows for a clearer delineation of the arc in the power spectrum.

In order to generate the secondary spectrum, $S_2(f_t,f_\lambda)$, we apply a Hamming window to the outer 10\% of the dynamic spectrum and compute its 2D Fourier transform, producing the amplitude spectrum. The squared magnitude of this then gives the secondary (or power) spectrum. We further shift and crop the spectrum to $f_\lambda > 0$ to remove the mirrored arc present for $f_\lambda < 0$. An example secondary spectrum is shown in the right panel of Figure \ref{fig:scint_obs}.

\section{Model}
\label{sec:model}
\begin{figure*}
    \centering
    \includegraphics[width=1\textwidth]{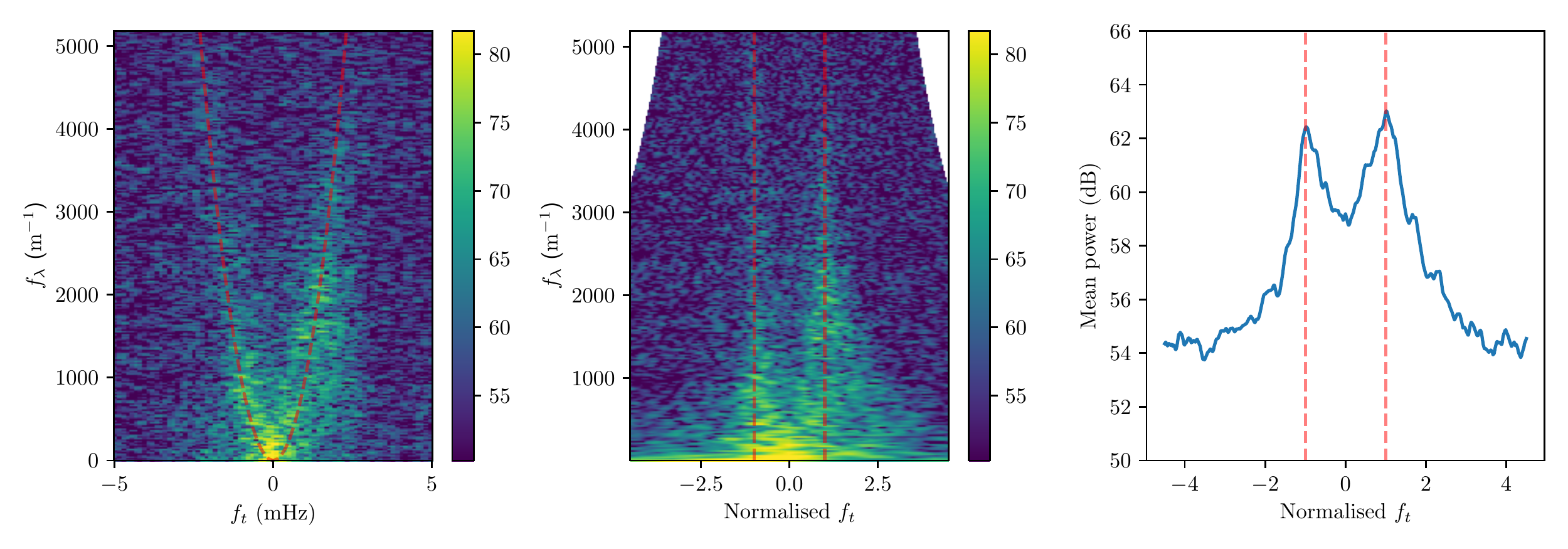}
    \caption{The secondary spectrum (left) is transformed so that parabolas become straight lines (middle), and a weighted sum along the $f_\lambda$ axis (see Equation \ref{eq:curvature_profile}) then gives the ``curvature profile" (right). In this way, each value along the horizontal axis of the curvature profile corresponds to a particular arc curvature, so distinct scintillation arcs manifest as peaks in the profile.}
    \label{fig:norm_sspec}
\end{figure*}
\subsection{Scintillation arc curvatures}
\label{sec:curvatures}
Physical models for the parabolic ``scintillation arcs" that appear in secondary spectra are discussed in detail in \citet{Walker+04} and \citet{Cordes+06}. We reproduce in this subsection the results important to our analysis.

Physical inhomogeneities in the ionized ISM diffractively scatter waves from a pulsar around the direct line of sight. The resulting phase differences between the incident light are described by some phase structure function $D_\phi(\boldsymbol{r})=\langle(\phi(\boldsymbol{r}^\prime)-\phi(\boldsymbol{r}^\prime+\boldsymbol{r}))^2\rangle$, where $\phi(\boldsymbol{r})$ is the phase of the signal originating from transverse sky position $\boldsymbol{r}$. The interference between these waves at the observer produces the scintillation effect. We can expect the ISM irregularities to follow a Kolmogorov power spectrum if they are turbulent in origin \citep{Armstrong+95}, for which the structure function takes the simple form $D_\phi(r)=(r/s_0)^{5/3}$, where $s_0$ is the diffractive spatial scale; see \citet{Cordes+06}. 

We use the Born variance to characterize the normalized root-mean-square intensity, or ``strength of scintillation," given by $m_B^2=0.773D_\phi(r_F)$ where $r_F$ is the Fresnel scale. In weak scintillation, defined by $m_\text{B}^2\lesssim 1$, the secondary spectra are effectively modelled by interference between a bright, lightly-scattered ``core" and a weak, scattered ``halo". Since the scintillation is weak, the self-scattering of the halo can be neglected. Consider a thin screen of ISM plasma localized at a distance $D_\text{psr-scr}$ from the pulsar. It is convenient to introduce the fractional distance $s = D_\text{psr-scr}/D$, where $D$ is the distance between the pulsar and Earth. Suppose that the screen scatters two wave components at angular sky positions $\boldsymbol{\theta}_1$ and $\boldsymbol{\theta}_2$, where $|\boldsymbol{\theta}_i| = \theta_i$ is the angle as measured from the pulsar's position $\boldsymbol{\theta}=\boldsymbol{0}$. The interference between these waves at the observer produces a single two-dimensional interference fringe pattern that varies slowly in phase with observing frequency. Observed as sinusoids in the dynamic spectrum, these fringes map to single Fourier components in the secondary spectrum. The positions of these components in the wavelength-resampled secondary spectrum $S_2(f_t, f_\lambda)$ are related to the scattering angles by
\begin{align}
	f_\lambda&=\frac{D(1-s)}{2s\lambda_\mathrm{c}^2}(\theta_2^2-\theta_1^2)\\
	f_t&=\frac{1}{s\lambda_\mathrm{c}}\boldsymbol{V}_\mathrm{eff}\cdot(\boldsymbol{\theta}_2-\boldsymbol{\theta}_1)
\end{align}
where $\lambda_\mathrm{c}=\nu_\mathrm{c}/c$ is the wavelength of the center frequency of the observation band, $c$ is the speed of light, and $\boldsymbol{V}_\text{eff}$ is the effective velocity (Equation \ref{eq:veff}). 

The angular dependence of the two Fourier coordinates suggests a quadratic relationship between them. In the case where one of the waves is undeflected (i.e. $\boldsymbol{\theta}_1=\boldsymbol{0}$) the relationship reduces to a simple parabola, $f_\lambda=\eta f_t^2$, with a curvature given by
\begin{align}\label{eq:eta}
    \eta=\frac{Ds(1-s)}{2(\boldsymbol{V}_\text{eff}\cdot\hat{\boldsymbol{\rho}})^2},
\end{align}
where $\hat{\boldsymbol{\rho}}=\boldsymbol{\theta}_2/\theta_2$ is the unit vector in the direction of the anisotropic component.

Weak scintillation produces a region of power in the secondary spectrum interior to the arc of maximum curvature, corresponding to $\boldsymbol{V}_\text{eff}\cdot\hat{\boldsymbol{\rho}}=V_\text{eff}$, which drops off rapidly outside of this parabola.
In the case of strong scintillation, the self-interference between different parts of the halo cannot be neglected. This results in the arc at the edge of the power losing contrast, as well as the appearance of inverted ``arclets" whose apexes lie along the main arc \citep{Brisken+09}. 

To determine the scintillation strength for our data, we use the relation $\nu_\text{c}/\Delta\nu_\text{d}=(r_\text{F}/s_0)^2$ involving the decorrelation bandwidth, $\Delta\nu_\text{d}$ \citep{Rickett+90}, to obtain $m_\text{B}^2 = 0.773(\nu_\text{c}/\Delta\nu_\text{d})^{5/6}$. Measurements of $\Delta\nu_\text{d}$ for our data (see middle panel of Figure \ref{fig:dm_variations}) give a typical $m_B^2$ on the order of 100, placing us well inside the strong scintillation regime. This is consistent with the broad arcs we see in the data, as well as the evidence for inverted arclets in some observations (see Section \ref{sec:discussion}).

\subsection{Normalized curvature profiles and probability distributions}
\label{sec:normsspec}
A useful technique for measuring arc curvatures, introduced in \citet{Reardon+20}, involves transforming the secondary spectrum in such a way that parabolas in $S_2(f_t, f_\lambda)$ map to straight lines. The resulting spectrum, known as the \textit{normalized secondary spectrum}, $\bar{S}_2(\bar{f}_t, f_\lambda)$ (see Figure \ref{fig:norm_sspec}), turns the integration of the power along the parabolas into a trivial integration along the $f_\lambda$ axis. The $\bar{f}_t$ coordinate is normalized to some reference curvature $\eta_0$ such that the curvature is related to the normalized Doppler variable by $\eta=\eta_0/\bar{f}_t^2$. From now on we refer to these $\bar{f}_t$ as ``normalized curvature," denoted by $\bar{\eta}=\sqrt{\eta_0/\eta}$. Per Equation \ref{eq:eta}, the normalized curvature is proportional to $\boldsymbol{V}_\text{eff}\cdot\hat{\boldsymbol{\rho}}$ in the anisotropic case, or just $V_\text{eff}$ for an isotropic screen.

Taking the (averaged) weighted integral of the 2D spectrum along $f_\lambda$ produces what we refer to as the \textit{curvature profile}:
\begin{align}\label{eq:curvature_profile}
	P(\bar{\eta})=\frac{1}{N_\lambda}\sum_{i=0}^{N_\lambda} \bar{S}_2(\bar{\eta}, f_\lambda^i)W(f_\lambda^i)
\end{align}
where $N_\lambda$ is the number of pixels along $f_\lambda$. The weight function $W(f_\lambda)\propto f_\lambda^{-7/3}$ describes the average power in the secondary spectrum along $f_\lambda$ for a Kolmogorov spectrum.

A number of our observations were passed through two separate digital filter banks (DFB3 and DFB4) and recorded in parallel. In these cases we simply take the linear average of the resulting pair of curvature profiles.

Since $\bar{\eta}$ and $-\bar{\eta}$ correspond to the same arc curvature, the two sides of the profile should be averaged. However, in the case of highly asymmetric profiles, averaging can significantly mute a distinct peak in power, often so much so that there is a new peak at an entirely different (sometimes infinite) curvature value. Such asymmetries can be understood from the interpretation of $f_t$ as a measure of the doppler shift due to the pulsar's line-of-sight motion, in which case an asymmetry between $\pm f_t$ implies a gradient in the signal about the line of sight \citep{Cordes+06}. These asymmetries can therefore provide information on the density and/or structure of the scattering screen, which is reflected in the power distribution of scattered images computed from the secondary spectra (see Section \ref{sec:interesting_spectra} for further discussion on generating and interpreting these scattered images). We find that most of our profiles are asymmetric and so we inspect the data visually and average only the profiles that are symmetric enough that there is no significant decrease in the prominence of the arc. For the rest of the observations we used an automated selection algorithm: for those that peak at $\bar{\eta}=0$ (i.e., $\eta=\infty$), we select the side with the highest integrated power as it is the stronger signal, while for the remaining profiles we simply select the side with the largest single power value. Observations that were extremely poor quality and/or featured obviously artificial regions of power were discarded. Of the 306 observations used in our analysis, 74 are classified as left-sided, 203 are classified as right-sided, 19 are averaged, and 10 are discarded. The classifications for all spectra are tabulated in the supplementary material referred to in Appendix \ref{sec:accessing_data}.

\begin{figure}[H]
    \centering
    \includegraphics[scale=0.35]{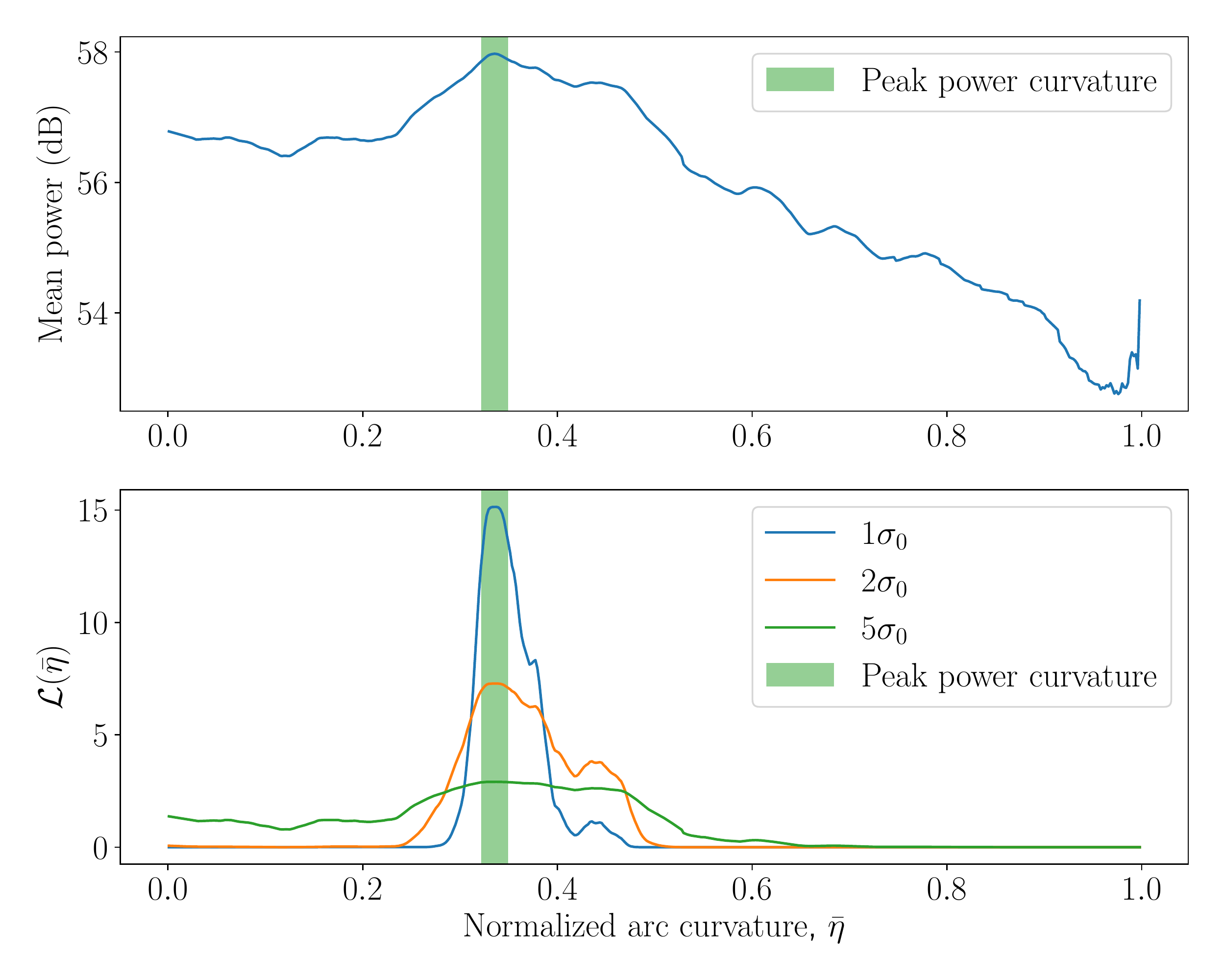}
    \caption{Averaged normalized curvature spectrum for observation at MJD 55864.04 (top) and examples of corresponding linear probability densities calculated using Equation \ref{eq:pdf} (bottom). The probability densities have been plotted using three different multiples of the standard deviation of the secondary spectrum $\sigma=\{\sigma_0,2\sigma_0,5\sigma_0\}$. The green bar shows the 68\% credible interval around the peak power, corresponding to the region in which the mean power is less than $1\sigma_0$ lower than the peak power. The choice of $\sigma$ appropriate for our data is discussed in Sections \ref{sec:method} and \ref{sec:static}.}
    \label{fig:norm_sspec_pdf}
\end{figure}

For an ideal, noise-free spectrum, the curvature of the scintillation arc corresponds to the peak power in the corresponding curvature profile. However, our method differs from previous analyses in that we do not use single measurements of the arc curvature for each observation. Instead we utilize the full curvature profiles by transforming them into probability densities (see Figure \ref{fig:norm_sspec_pdf}), which we take as the likelihood functions $\mathcal{L}$ when modelling the time variations (described in Section \ref{sec:method}). This has three distinct advantages over the standard approach:
\begin{enumerate}
\item It allows us to use observations with extremely high curvatures, where the opposite sides of the arc merge and the peak can no longer be reliably identified.

\item It does not require any assumptions about the form of $\mathcal{L}$ for the arc curvature variations (though we must still assume some form for the noise in the individual profiles).

\item It allows for contributions from additional arcs that would otherwise be ignored by a single arc measurement. This is important because the arc with the highest power may not correspond to the same scattering screen as the majority of the other observations, but rather a different screen that temporarily contributes a stronger signal.
\end{enumerate}

We calculate the probability distribution corresponding to the curvature profile $P(S_i, \bar{\eta})$ of spectrum $S_i$ by taking some Gaussian noise value $\sigma$ and assigning a probability to the normalized curvatures based on their deviation from the peak power $P_{\text{max}}$ of the profile:
\begin{align}\label{eq:pdf}
   \ln {\cal L}(S_i | \bar{\eta})=-\frac{1}{2}\left(\frac{P(\bar{\eta},S_i)-P_\text{max}(S_i)}{\sigma(S_i)}\right)^2+\ln N
\end{align}
where $N$ is a suitable normalizing factor. The noise value $\sigma$ would nominally be taken to be the Gaussian noise of the secondary spectrum, however we expect other sources of noise to be present in our data, such as inherent scatter in the arc curvature due to short-timescale velocity and anisotropy variations. The specification of $\sigma$ is detailed in Sections \ref{sec:method} and \ref{sec:static}.

We truncate the probability density function at the chosen reference curvature $\eta_0$, which must be small enough compared to the observed curvatures that we can reasonably assume the probability at $\eta<\eta_0$ to be negligible. Our observations are all measured to have $\eta$ of $\gtrsim 10^3\text{ m}^{-1}\text{mHz}^{-2}$, so we choose $\eta_0=10^2\text{ m}^{-1}\text{mHz}^{-2}$.

\subsection{Velocity model}
\label{sec:velocity}
Per Equation \ref{eq:eta}, the arc curvature variations are determined by variations in the fractional screen distance, $s$, anisotropy orientation $\hat{\boldsymbol{\rho}}$, and the effective velocity, $\boldsymbol{V}_\text{eff}$. The effective velocity is the velocity of the point on the screen intersected by the line-of-sight between the Earth and the pulsar in the frame of the ISM. It is given by \citep{Cordes+98}
\begin{align}\label{eq:veff}
    \boldsymbol{V}_\text{eff}=(1-s)(\boldsymbol{V}_\text{p}+\boldsymbol{V}_\mu)+s\boldsymbol{V}_\text{E}-\boldsymbol{V}_\text{ISM} ,
\end{align}
where $\boldsymbol{V}_\text{p}+\boldsymbol{V}_\mu$ is the velocity of the pulsar decomposed into its orbital and transverse (proper motion) parts, respectively, $\boldsymbol{V}_\text{E}$ is the Earth velocity, and $\boldsymbol{V}_\text{ISM}$ is the velocity of the screen. We take the reference frame to be that of the Solar System barycentre. The proper motion has been measured to high precision and Earth's velocity is known. The pulsar orbital motion can be decomposed into components parallel and perpendicular to the line of nodes. As a function of orbital phase $\phi=\theta+\omega$, with $\theta$ the true anomaly and $\omega$ the longitude of periastron, these are given by
\begin{align}
	V_{\text{p},\parallel}&=-V_0(e\sin\omega+\sin\phi)\\
	V_{\text{p},\perp}&=V_0\cos i(e\cos\omega+\cos\phi)
\end{align}
where $i$ is the inclination angle, $e$ is the eccentricity of the orbit, and
\begin{align}
	V_0=\frac{2\pi xc}{P_\text{b}\sin i\sqrt{1-e^2}}\approx\frac{23.8\text{ km s}^{-1}}{\sin i}
\end{align}
is the mean orbital velocity in terms of the projected semi-major axis, $x$ (in units of time) and orbital period $P_\text{b}$. The components of $\boldsymbol{V}_\text{p}$ are rotated by an angle $\Omega$, known as the longitude of ascending node, to obtain the RA and DEC components, $V_{\text{p},\alpha},V_{\text{p},\delta}$. The only parameters that are not precisely known are $i$ and $\Omega$, which only have weak constraints from pulsar timing (see Section \ref{sec:timing_comparison}). We therefore treat these as free parameters.

We take the anisotropy orientation to be $\hat{\boldsymbol{\rho}}=(\sin\psi,\,\cos\psi)$ and fit for $\psi$, the angle of the anisotropy as measured east from the declination axis. Finally, the ISM velocity is not known \textit{a priori} and so we treat it as a free parameter. For an isotropic screen, $\psi$ does not play a role and we simply have $\eta\propto 1/V_\text{eff}^2$ for the curvature of the outer edge of the power, in which case we use ISM velocity components $V_{\text{ISM},\alpha}$ and $V_{\text{ISM},\delta}$. However, for an anisotropic screen, only the component of the ISM velocity along the direction of anisotropy, $\boldsymbol{V}_\text{ISM}\cdot\hat{\boldsymbol{\rho}}$, contributes. We therefore take this to be the single ISM velocity parameter, $V_{\text{ISM},\psi}$.

Although we expect the pulsar orbital parameters and distance to remain constant over the full timespan of the data, due to the motions and inhomogeneity of the ISM there is no reason to expect the scattering to be dominated by the same ISM feature throughout. Indeed, the large variations in DM shown in Figure \ref{fig:dm_variations} may be the result of independent scattering screens, in which case single $s$, $\psi$, and $V_\text{ISM}$ parameters will not accurately model the data. For this reason we also fit models with several $s$, $\psi$, and $V_\text{ISM}$ parameters, each applied to a distinct subset of the data---which we refer to as ``epochs"---so as to capture potential time variability in the dominant source of scattering. Since we do not know the nature of the time variation, the boundaries of the epochs are also taken to be free parameters in the model.

\section{Method}
\label{sec:method}
\begin{figure*}
    \centering
    \includegraphics[scale=0.44]{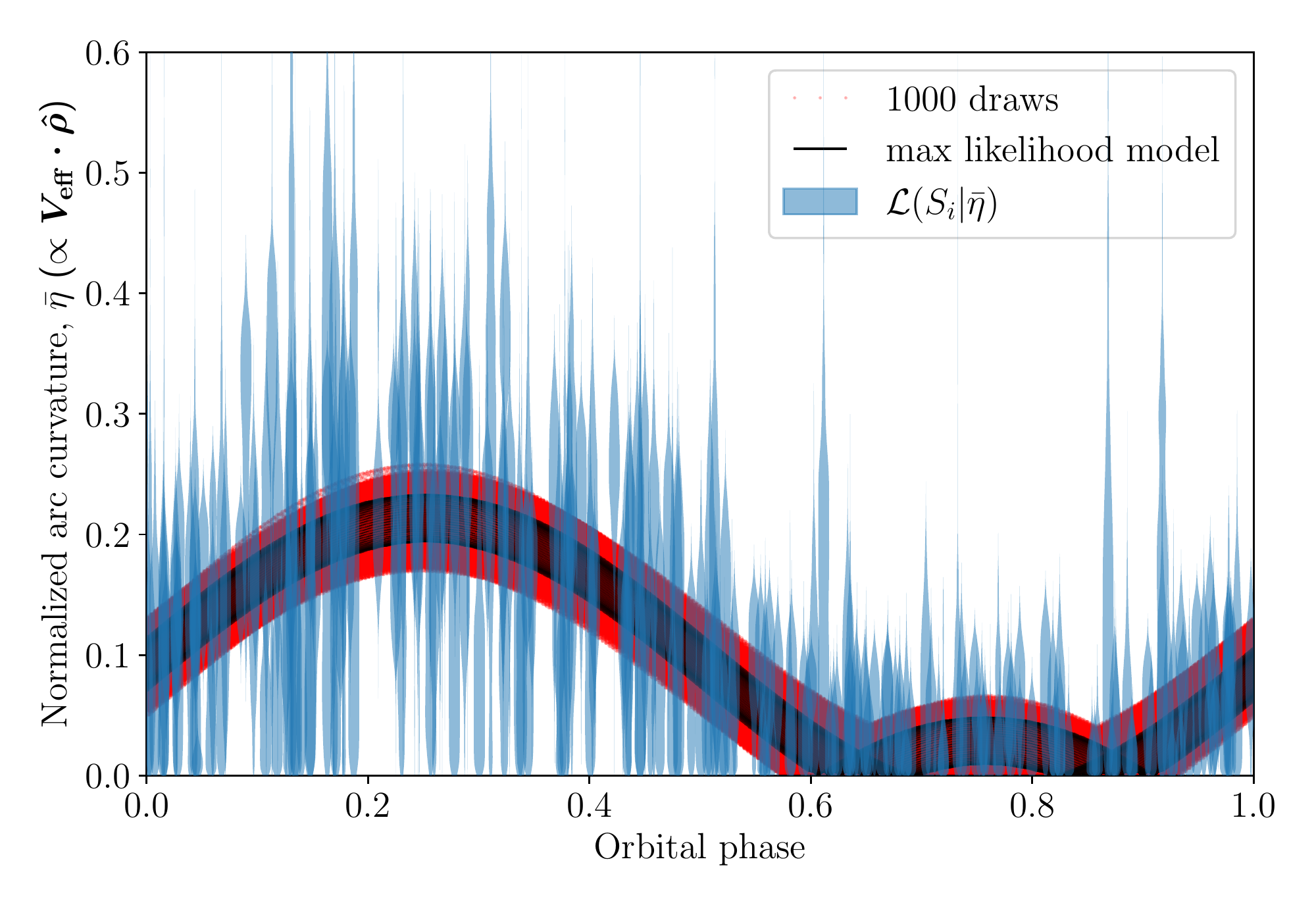}
    \includegraphics[scale=0.44]{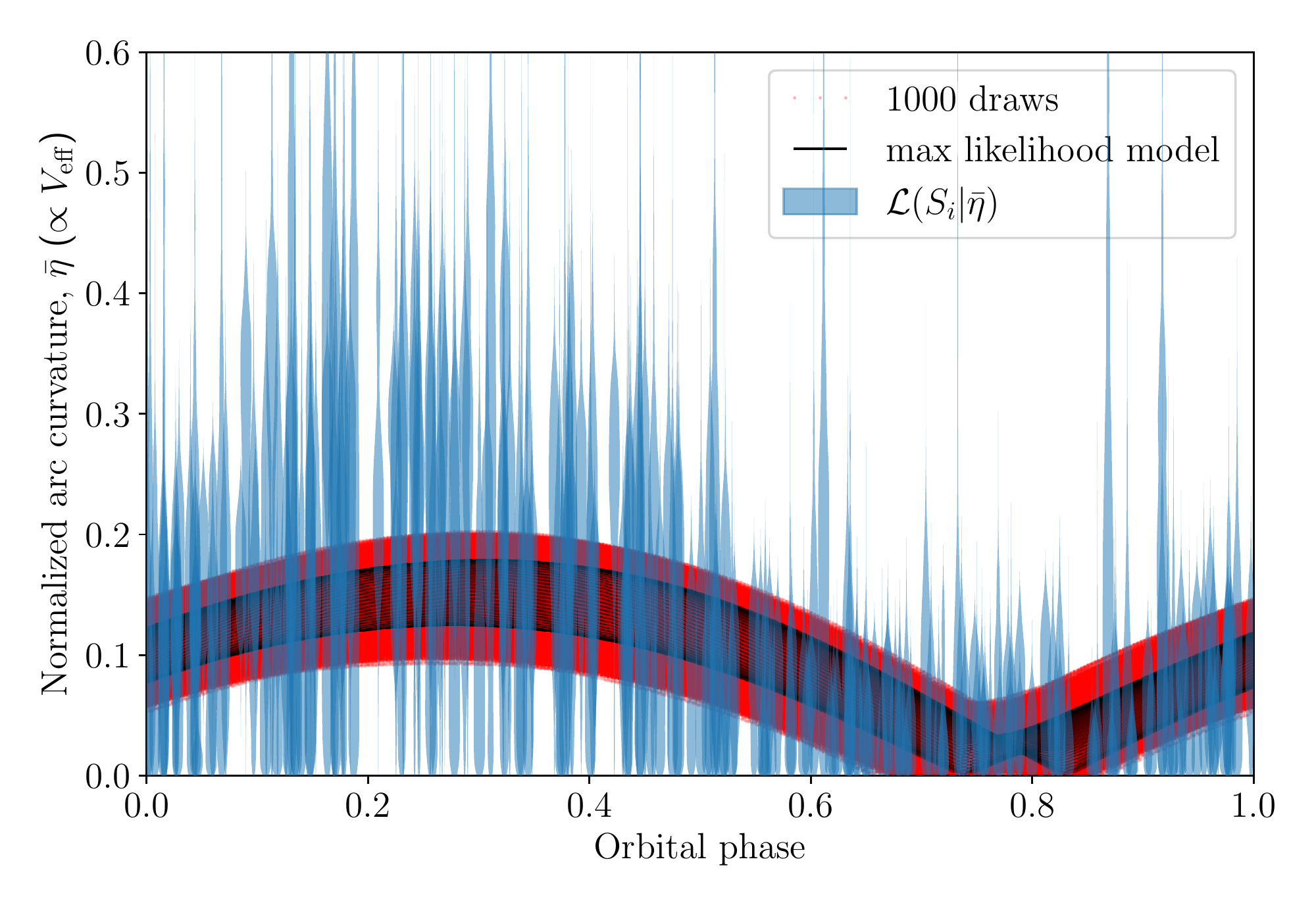}
    \caption{Plots of the data and model with respect to orbital phase for an anisotropic (left) and isotropic (right) screen. Since the measurements used in our analysis do not come in the form of data points and associated uncertainties, we use in their place violin plots (shown in blue) with widths proportional to the probability distributions ${\cal L}(S_i | \bar{\eta})$. The noise values (see Equation \ref{eq:sigma}) used in the calculation of these ``data points" use the highest-likelihood white noise ($\alpha$, $\beta$) parameters. The black ``regions" are actually each a single curve calculated from the curvature model using the maximum likelihood parameters. The non-zero width spanned by the maximum likelihood curve is a result of long-term variations induced by the Earth's orbital motion. The red corresponds to the region spanned by 1,000 model curves using parameters sampled randomly from the posterior distribution and therefore defines a 99.9\% confidence region around the maximum likelihood model.}
    \label{fig:model_fit}
\end{figure*}
We use Bayesian inference to fit the velocity model to the arc curvatures. For an introduction to Bayesian methods, with examples drawn from gravitational-wave astronomy, see \citet{Thrane+19}.
As described in Section \ref{sec:normsspec}, we take the likelihood, ${\cal L}(S_i | \bar{\eta})$, for each observation, $S_i$, of J1603-7202 to be the probability distribution calculated from the corresponding normalized curvature profiles $P_i(\bar{\eta})$ using Equation \ref{eq:pdf}. The likelihood for the full dataset $\{S\}$ given the parameters $\theta$ is then
\begin{align}
    {\cal L}(\{S\} | \theta) = & \prod_i {\cal L}(S_i | \theta) \\
    = & \prod_i \int \mathrm{d}\bar{\eta} \, {\cal L}(S_i | \bar{\eta}) \pi(\bar{\eta}_i | \theta) \\
    = & \prod_i \int \mathrm{d}\bar{\eta} \, {\cal L}(S_i | \bar{\eta})\,\delta\big(\bar{\eta} - \bar{\eta}_i(\theta)\big)\\
    = & \prod_i {\cal L}(S_i | \bar{\eta}_i(\theta))
 \end{align}
where $\pi(\bar{\eta}_i | \theta)=\delta\big(\bar{\eta} - \bar{\eta}_i(\theta)\big)$ is the (trivial) prior for the normalized curvature $\bar{\eta}$ at observation $i$ given the parameters $\theta$. The posterior distribution is then given by
\begin{align}
    p(\theta|\{S\})=\frac{{\cal L}(\{S\} | \theta)\pi(\theta)}{Z}
\end{align}
where $\pi(\theta)$ is the prior on the parameters (see Section \ref{sec:priors}) and $Z=\int {\cal L}(\{S\}|\theta)\,\pi(\theta)\,\mathrm{d}\theta$ is the Bayesian evidence.

The noise $\sigma$ for each observation is adjusted alongside the physical model parameters through $\alpha$ and $\beta$ white noise modifiers commonly adopted in pulsar timing applications (and referred to as ``EFAC" and ``EQUAD" respectively):
\begin{align}\label{eq:sigma}
    \sigma^2 = (\alpha\,\sigma_0)^2+\beta^2
\end{align}
where $\sigma_0$ is simply the standard deviation of the secondary spectrum measured away from the arc and acts as an initial guess for the noise. The white noise modifiers are intended to absorb extra contributions to scatter that cannot be easily modelled, such as from stochastic changes in the scattering plasma through short-timescale velocity and anisotropy variations, or a poor estimate of the secondary spectrum noise. We discuss further whether this prescription is appropriate for our data in Section \ref{sec:static}.

\subsection{Priors}
\label{sec:priors}
Our choices for the priors $\pi(\theta_i)$ for the free parameters in our model are designed to be conservative. The fractional screen distance, $s$, and anisotropy angle, $\psi$, are totally unconstrained (though we expect from the lack of obvious annual curvature modulations that the screen is relatively close to the pulsar so that $s\lesssim 0.5$) and we adopt uniform priors for them. Though the longitude of ascending node, $\Omega$, has constraints from pulsar timing, they are too weak to be useful for this analysis, so we adopt a uniform distribution for $\Omega$. We assume a uniform distribution in $\cos i$ to reflect the on-sky distribution of binary inclination angles. For the ISM velocities, $V_{\text{ISM},\alpha}$, $V_{\text{ISM},\delta}$, and $V_{\text{ISM},\psi}$, we adopt normal distribution priors centered on $0\text{ km s}^{-1}$ with a width of $100\text{ km s}^{-1}$. A distance to the pulsar of $D=3.4\pm 0.5$ kpc has recently been reported in \citet{Reardon+21}, which we take as a Gaussian prior. Finally, the noise parameter $\beta$ is taken to be uniformly distributed while the $\alpha$ parameter is taken to be log-uniformly distributed.

We carry out the inference calculation using the \textsc{dynesty} dynamic nested sampling routines included in the \textsc{bilby} inference library \citep{Ashton+19}.

\section{Results}
\label{sec:results}
We fit several different models to the data: an isotropic and an anisotropic screen model with all parameters fixed (one epoch) and time-varying anisotropic screen models with two to five epochs. The results from the static models are detailed in Section \ref{sec:static} while those from the multiple-epoch models are given in Section \ref{sec:tvary}.

To assess the relative statistical significance of the different models we use the (log) Bayes factor $\ln B = \ln Z_1 - \ln Z_2$, which is the ratio of the evidences for two different models. It should be noted however that we do not have precise knowledge of the noise in the secondary spectra, which we are assuming to be Gaussian. The presence of non-Gaussian contributions to the noise, may mean the noise model is misspecified. The precise numerical values for the Bayes factors may therefore not be completely reliable, so we only use them conservatively as a qualitative comparison tool.

\subsection{Static models}
\label{sec:static}
For the static model, we find that the anisotropic case is highly favored, with a log Bayes factor of $\ln B\approx 57$ and a visually better fit to the data. In Figure \ref{fig:model_fit} we plot the predicted $\bar{\eta}$ as a function of orbital phase for both the anisotropic (left) and isotropic (right) models alongside the observations. Since our analysis does not use single curvature measurements, we instead display the data as violin plots of the probability distributions $\mathcal{L}(S_i|\bar{\eta})$. The black ``regions" are actually each a single curve corresponding to the highest-likelihood model parameters, while the red bands are the regions spanned by 1,000 models with parameters sampled randomly from the posterior distribution. The non-zero width spanned by the maximum-likelihood model curve is a result of long-timescale variations caused by the Earth's orbital motion. There is a unique minimum in the isotropic model curve corresponding to the minimum effective velocity in the orbit, while the anisotropic model features a pair of minima, corresponding to the points at which the effective velocity becomes perpendicular to the anisotropy: $\boldsymbol{V}_\text{eff}\cdot\hat{\boldsymbol{\rho}}=0$.

The presence of anisotropic structure has been inferred in a number of prior studies \citep{Trang+07,Brisken+09,Stinebring+19,Sprenger+20,Rickett+21,McKee+22}, our result therefore provides additional credence to the idea that elongated plasma structure is a common feature in the ISM, although our model is not sensitive to the degree of anisotropy.

The parameter values  from the fit are shown in Table \ref{tab:params} and a corner plot of marginal posterior distributions is shown in Figure \ref{fig:corner}. The anisotropic model does not significantly favour either one of the two degenerate solutions in $\cos i$ and $\Omega$: $i=25^\circ\substack{+6^\circ \\ -4^\circ},\, \Omega=132^\circ\substack{+11^\circ \\ -13^\circ}$ and $i=155^\circ\substack{+5^\circ \\ -7^\circ},\,\Omega=327^\circ\pm 14^\circ$.

\begin{table}[h]
    \centering
    \caption{Parameter fit values with $1\sigma$ credible intervals. Where more than one value is shown there is a degeneracy in the posterior for that parameter. The $V_\text{ISM}$ values are the estimates after subtracting out the contribution from differential rotation.}
    \begin{tabular}{cccc}
		\toprule
		Parameter & Isotropic & Anisotropic & Anisotropic \\
		& & 1 epochs & 3 epochs \\
		\midrule
		$\cos i$ & $0.91\substack{+0.02 \\ -0.05}$ & $0.91\substack{+0.03 \\ -0.05}$ & $0.86\substack{+0.03 \\ -0.06}$ \\
		& & $-0.92\substack{+0.06 \\ -0.03}$ & $-0.85\substack{+0.06 \\ -0.04}$ \\
        $\Omega$ ($^\circ$) & $111 \pm 11$ & $132\substack{+11 \\ -13}$ & $110\substack{+15 \\ -11}$ \\
        & & $327\pm 14$ & $305\substack{+19 \\ -12}$ \\
        $s$ & $0.47\pm 0.06$ & $0.25\substack{+0.09 \\ -0.08}$ & See Table \ref{tab:tvary_params} \\
        $D$ (kpc) & $4.0 \pm 0.4$ & $3.3 \pm 0.5$ & $3.7 \pm 0.4$ \\
        $V_{\text{ISM},\alpha}$ (km s$^{-1}$) & $21\pm 6$ & - & - \\
        $V_{\text{ISM},\delta}$ (km s$^{-1}$) & $-44\pm 10$ & - & - \\
        $V_{\text{ISM},\psi}$ (km s$^{-1}$) & - & $-5\substack{+15 \\ -18}$ & See Table \ref{tab:tvary_params} \\
        $\psi$ ($^\circ$) & - & $50 \pm 13$ & See Table \ref{tab:tvary_params} \\
        $\ln(\alpha/1\,\mathrm{dB})$ & $1.4\substack{+0.8 \\ -1.7}$ & $1.2\substack{+0.7 \\ -1.7}$ & $0.8\substack{+0.8 \\ -1.5}$ \\
        $\beta$ (dB) & $1.8\substack{+0.2 \\ -0.7}$ & $1.2\substack{+0.2 \\ -0.6}$ & $1.0\substack{+0.1 \\ -0.4}$ \\        
		\bottomrule
    \end{tabular}
    \label{tab:params}
\end{table}
We find that the scattering screen is closer to the pulsar than to the Earth, at a fractional distance of $s=0.25\substack{+0.09 \\ -0.08}$. This is consistent with the lack of obvious annual modulation in the data, as a small $s$ diminishes the $s\boldsymbol{V}_\text{E}$ term in the effective velocity.

\begin{figure}[H]
    \centering
    \includegraphics[scale=0.6]{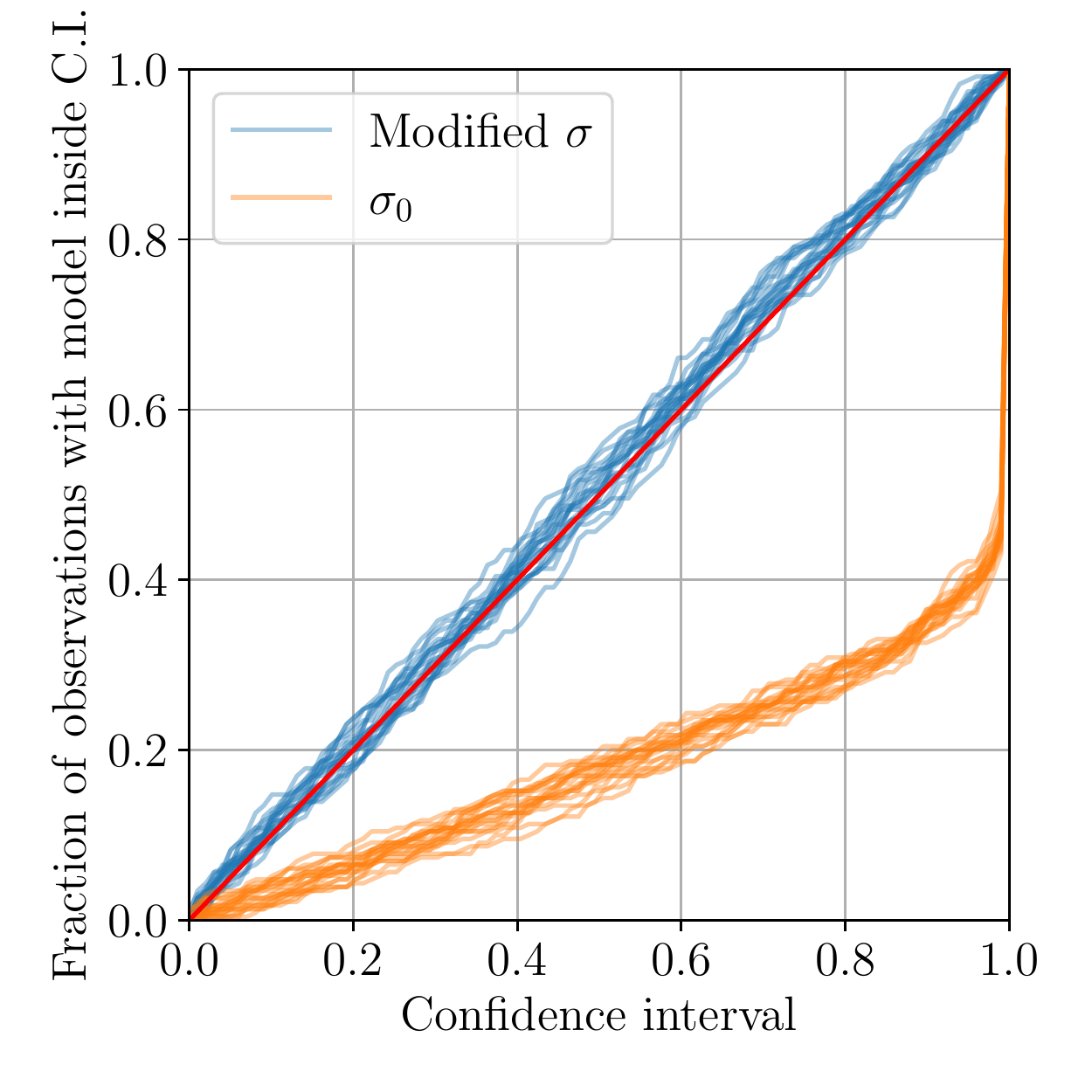}
    \caption{Parameter-parameter (pp) plot for the static (single-epoch) model, which displays the fraction of model predictions that lie within a given confidence interval of their corresponding $\mathcal{L}(S_j|\eta)$. Thus, for a perfectly specified noise model, the resulting graph would be a straight line of slope 1 (shown in red). In blue is the plot generated using the modified noise (Equation \ref{eq:sigma}) to calculate $\mathcal{L}(S_j|\eta)$, while the orange uses the unmodified noise, $\sigma_0$. For each of these two cases we define 95\% confidence regions by plotting 20 curves, each corresponding to a distinct set of parameters sampled randomly from the posterior distribution. The curves using the unmodified noise are highly inconsistent with the diagonal, suggesting a significant amount of unmodelled scatter. The noise modification absorbs this scatter and the resulting pp curves closely follow the diagonal, suggesting that the uncertainties for our parameter measurements are robust.}
    \label{fig:pp_plot}
\end{figure}

As mentioned in Section \ref{sec:method}, the reliability of the model fit and reported parameter uncertainties is dependent on the validity of the white noise modification. To assess this we generate a parameter-parameter (pp) plot \citep{Cook+06}, which displays the fraction of observations for which the maximum likelihood model lies within a given confidence interval. That is, for each observation $j$ we determine whether the maximum-likelihood model prediction $\eta_j(\theta_\text{max})$ lies within a particular confidence interval of $\mathcal{L}(S_j|\eta)$. The pp-plot then displays the fraction of model points that lie within the interval, against the confidence interval itself expressed as a fraction of 1. Thus, for a perfectly specified noise model, the resulting graph would be a straight line of slope 1. The pp-plots for the static anisotropic model using the unmodified noise, $\sigma_0$ (orange), and modified noise, Equation \ref{eq:sigma} (blue), are shown in Figure \ref{fig:pp_plot}. The curves for the unmodified noise are highly inconsistent with the $\text{slope}=1$ line shown in red, while the modified noise follows the diagonal closely. This implies there is a relatively large amount of unmodelled scatter in the data that is being absorbed by the white noise modifiers. The parameter measurements should therefore be interpreted as the ``average" of the variations in the parameter values causing the scatter. Given the agreement between the pp-plot and the diagonal, we expect the associated parameter uncertainties to be robust.

\subsection{Time-varying models}
\label{sec:tvary}
The Bayes factors of the multi-epoch models relative to the anisotropic single epoch model discussed above are given in Table \ref{tab:tvary_bayes}. We find weak support for two epochs and moderately strong support for three or more epochs, with the Bayes factors changing only negligibly for more than three epochs.

\begin{figure}[H]
    \centering
    \includegraphics[scale=0.4]{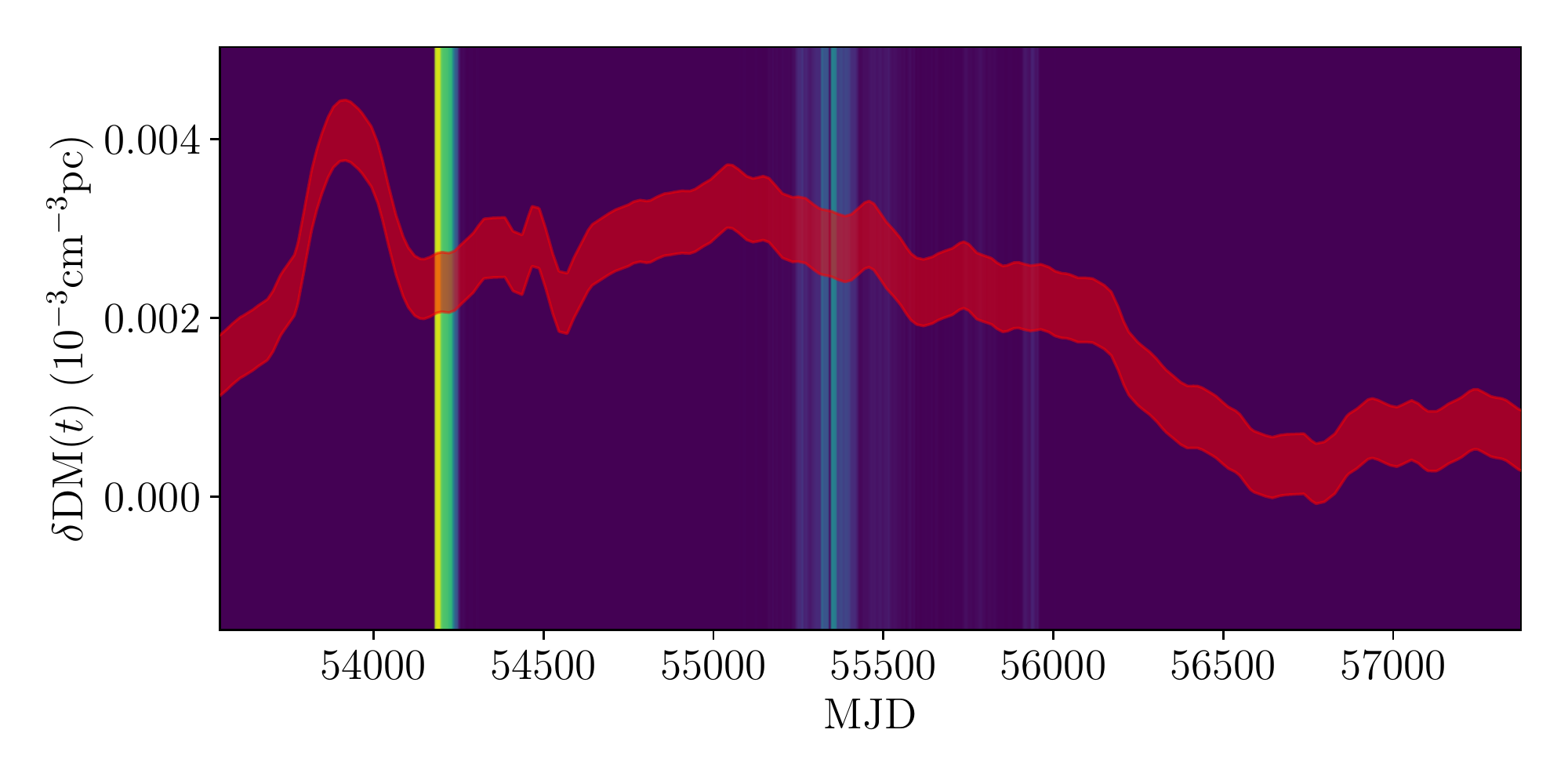}
    \caption{1D posteriors of the epoch boundaries overlaid on a plot of the dispersion measure variations. The first epoch spans the period over which the ESE spike occurred.}
    \label{fig:dm_epoch}
\end{figure}

\begin{table}[H]
    \centering
    \caption{Bayes factors of the anisotropic multiple-epoch models relative to the anisotropic single-epoch model.}
    \begin{tabular}{cc}
        \toprule
    	Anisotropic model & $\ln B$ \\
    	\midrule
    	1 epoch & 0.0 \\
        2 epochs & 4.2 \\
        3 epochs & 13.2 \\
        4 epochs & 12.8 \\
        5 epochs & 13.5 \\
        \bottomrule
    \end{tabular}
    \label{tab:tvary_bayes}
\end{table}
The values of the fixed parameters for the three-epoch model are given in Table \ref{tab:params}, while the values of the time varying parameters are given in Table \ref{tab:tvary_params}. Figure \ref{fig:dm_epoch} shows the dispersion measure variations overlaid on the 1D marginalized posteriors for the boundaries of the epochs. A comparison between the marginalized posteriors of the fixed parameters for the one- and three-epoch models is shown in Figure \ref{fig:tvary_corner}. The fixed parameters do not change substantially between the one- and three-epoch models, with the greatest change being the decrease in $\Omega$. The $\alpha$ and $\beta$ white noise modifiers are lower compared to those of the single-epoch model, suggesting that the modifiers for the single-epoch model may be absorbing some time-variability in the ISM parameters.

\begin{table}[H]
    \centering
    \caption{Fit values for the three time varying parameters as well as the day corresponding to the boundary between neighbouring epochs. The $V_\text{ISM}$ values are the estimates after subtracting out the contribution from differential rotation.}
    \begin{tabular}{ccccc}
        \toprule
    	Epoch & $s$ & $\psi$ ($^\circ$) & $V_{\text{ISM},\psi}$ & Epoch end \\
    	& & & (km s$^{-1}$) & ($\text{MJD} -$\\
    	& & & & $53548.363$)\\
    	\midrule
        1 & $0.32\pm 0.09$ & $29\substack{+17 \\ -16}$ & $-30\substack{+15 \\ -17}$ & $664\substack{+23 \\ -20}$ \\
        2 & $0.10\substack{+0.04 \\ -0.03}$ & $35\substack{+20 \\ -16}$ & $-35\substack{+20 \\ -17}$ & $1820\substack{+182 \\ -67}$ \\
        3 & $0.15\substack{+0.06 \\ -0.05}$ & $33\substack{+18 \\ -17}$ & $-30\substack{+17 \\ -18}$ & $3828.492$\\
        \bottomrule
    \end{tabular}
    \label{tab:tvary_params}
\end{table}

\section{Discussion}
\label{sec:discussion}
\subsection{Comparison with pulsar timing}
\label{sec:timing_comparison}
\begin{figure*}
    \centering
    \includegraphics[scale=0.62]{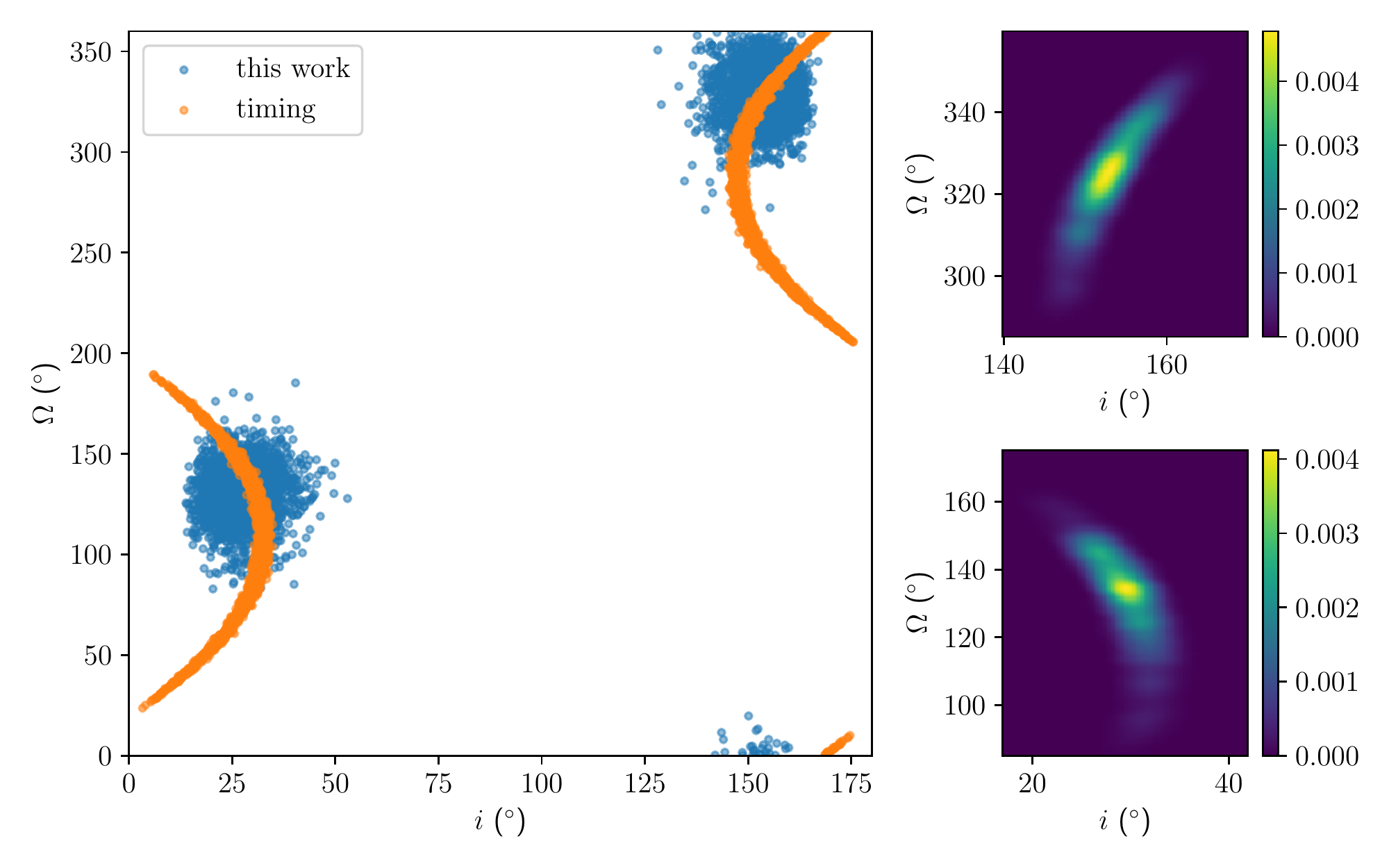}
    \caption{Left: comparison between the $i$-$\Omega$ posterior samples for the one-epoch model (blue) and those from pulsar timing (orange), right: the result from combining kernel density estimates of the two posteriors. Our scintillation analysis is an improvement over the constraints from timing alone, particularly for $\Omega$. However, there remains the same degeneracy in the solutions.}
    \label{fig:timing_comparison}
\end{figure*}
The degenerate $i$-$\Omega$ marginalized posteriors are consistent with results from pulsar timing \citep{Reardon+21} for both the one- and three-epoch models. The one-epoch model posterior and the timing posterior are shown overlaid in the left panel of Figure \ref{fig:timing_comparison}. Though the timing posterior does not precisely constrain $i$ or $\Omega$ individually, its distribution in $i$-$\Omega$ space is narrow and so provides extra verification for our model. By multiplying kernel density estimates of the timing and scintillation likelihoods, we obtain the posterior distributions shown in the right panel of Figure \ref{fig:timing_comparison} and further improved constraints: $i=31^\circ\substack{+3^\circ \\ -2^\circ},\,\Omega=135^\circ\substack{+9^\circ \\ -12^\circ}$ and $i=153^\circ\pm 3^\circ,\,\Omega=327^\circ\pm 9^\circ$.

\subsection{Screen properties}
The single-epoch (static) model gives the screen distance as $D_s = (1 - s)D=(2.5\pm 0.5)\text{ kpc}$. We do not find any convincing coincidence between this distance and those of stars in the vicinity of J1603$-$7202's sky location that may be the source of the extreme scattering event. The only catalogued star with a distance consistent with that of the scattering screen is Gaia DR2 5806675731270113280, however it has a substantial distance uncertainty of 1 kpc, making the association dubious. Furthermore, the star's velocity along the anisotropy direction is significantly higher than our measured $V_{\text{ISM},\psi}$ \citep{Gaia+18}. 

The measured component of the ISM velocity along the direction of the anisotropy for the static model is $-47\substack{+15 \\ -18}\text{ km s}^{-1}$. At a distance on the order of a kiloparsec, contribution from the differential rotation of the galaxy may significantly affect this value. We estimate this velocity by taking the orbital velocities of the Earth and the screen about the Galactic center to be equal at $V_c=220$\,km\,s$^{-1}$ \citep{Majewski+07} \citep[i.e. assuming a flat rotation curve; see][]{Reid+14}. From this we obtain a differential velocity component along the direction of the anisotropy of $\sim -40\text{ km s}^{-1}$, suggesting that the vast majority of $V_{\text{ISM},\psi}$ is accounted for by differential rotation. The remaining $\sim 10\text{ km s}^{-1}$ is consistent with the thermal or Alfv\'{e}n speed of the interstellar plasma \citep{Goldreich+95}.

The fit values of the screen parameters for the three-epoch model, shown in Table \ref{tab:tvary_params}, suggest that the scattering was initially dominated by a screen at $s=0.32\pm 0.09$ before transitioning to a screen closer to the pulsar. The size of the uncertainties makes it unclear whether epoch 2 and epoch 3 truly correspond to separate screens or not. Like with the static model, there is no convincing coincidence between these distances and those of stars around the line-of-sight to J1603$-$7202.

It is interesting to note that the first epoch coincides with the ``spike" of the extreme scattering event (see Figure \ref{fig:dm_epoch}), suggesting that the spike in DM was the result of a separate, transient screen located slightly closer to the Earth.

However, the fact that $\psi$, and to an extent $V_{\text{ISM},\psi}$, have such similar values between the epochs may cast doubt on this interpretation. Naively, we would expect independent screens to show no correlation between the $s$, $\psi$, and $V_{\text{ISM},\psi}$ parameters, so the fact that we see this may be a sign that the varying $s$ values are over-fitting noise in the data and do not correspond to genuine screens.

\subsection{Companion mass and Shapiro delay}
\begin{figure*}
    \centering
    \includegraphics[scale=0.7]{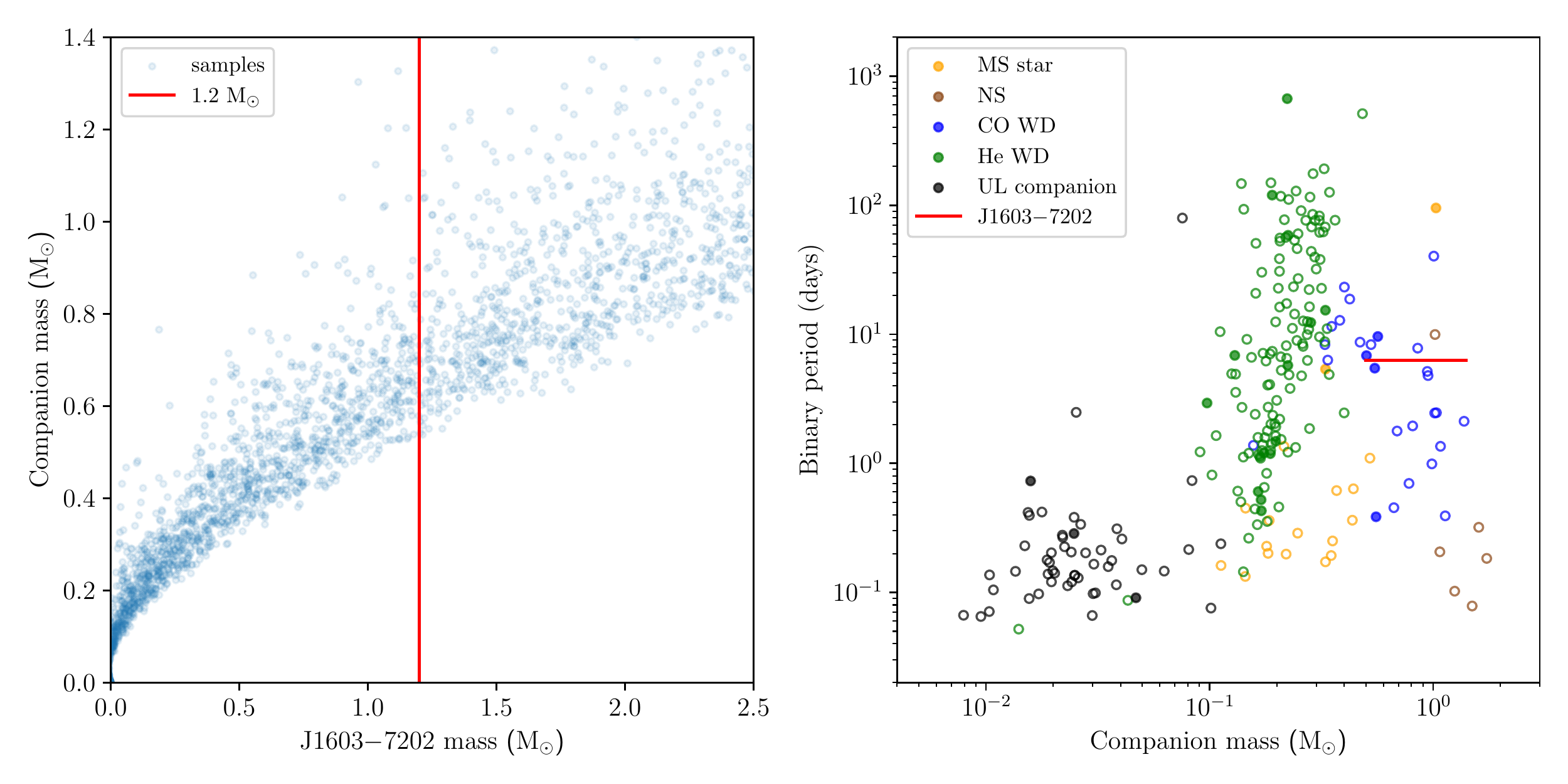}
    \caption{Samples from the posterior distribution for $m_\text{p}$-$m_\text{c}$ (left), calculated using Equation \ref{eq:mass_function} and a uniform distribution in the companion mass. Assuming a lower bound on J1603$-$7202's mass of $\sim 1.2\,\text{M}_\odot$ (vertical red line) gives a lower bound on the companion mass of $\sim 0.5\,\text{M}_\odot$. The plot on the right shows the mass-period distribution of companions for pulsars with pulse periods of $<30\text{ ms}$. The companion types are color-coded by type: main-sequence (MS) stars (orange), neutron stars (NS) (brown), carbon-oxygen (CO) and helium (He) white dwarfs (blue and green, respectively), and ultra-light companions or planets (black). The unfilled points correspond to those that only have a median mass estimate (assuming $i=60^\circ$ and $m_\text{p}=1.35\,\text{M}_\odot$) available. The range for J1603$-$7202's companion is shown in red and coincides with the region populated by CO white dwarfs.}
    \label{fig:companion_mass}
\end{figure*}
Precise measurements of the projected semi-major axis $x$ and period $P_\text{b}$ of J1603$-$7202's orbit \citep{Reardon+21} gives a value for the binary mass function of
\begin{align}\label{eq:mass_function}
    f(m_\text{p},m_\text{c})=\frac{(m_\text{c}\sin i)^3}{(m_\text{p}+m_\text{c})^2}=\frac{4\pi^2c^3}{G}\frac{x^3}{P_\text{b}^2}= 0.0087879\,\mathrm{M}_\odot,
\end{align}
where $G$ is the gravitational constant.

The mass function allows us to place a plausible lower bound on the mass of J1603$-$7202's companion given the observed distribution of binary pulsar masses. Figure \ref{fig:companion_mass} shows samples from the posterior distribution in $m_\text{p}$-$m_\text{c}$ assuming a uniform distribution in the companion mass and calculating $m_\text{p}$ using Equation \ref{eq:mass_function}. To date, the smallest known (recycled) binary pulsar masses are $\sim 1.2\,\text{M}_\odot$ \citep{Ozel+16}. Taking this as the lower bound for J1603$-$7202 gives a lower bound on the companion mass of $\sim 0.5\,\text{M}_\odot$. The right plot of Figure \ref{fig:companion_mass} shows the distribution of pulsar companions in $m_\text{c}$-$P_\text{b}$ space. The region corresponding to J1603$-$7202's companion is marked in red and seems to rule out a He white dwarf, instead coinciding with the region populated by CO white dwarfs \citep{Manchester+05}. Given J1603$-$7202's very low eccentricity of $\sim 9.32\times 10^{-6}$, this would seem to suggest evolution via steady Roche lobe overflow from the donor star onto J1603$-$7202 \citep{Tauris+12}.

This lower bound on the companion mass also gives us a lower bound on the peak Shapiro delay, which is given by
\begin{align}\label{eq:shapiro_delay}
    \Delta_\text{Shapiro}=-\frac{2G}{c^3}m_\text{c}\ln(1-\sin i).
\end{align}
Similarly to the pulsar mass, we generated a posterior distribution in $\Delta_\text{Shapiro}$-$m_\text{c}$ space using a uniform companion mass distribution. For a companion mass $\gtrsim 0.5\,\text{M}_\odot$, we get a Shapiro delay of $\gtrsim 1.6\,\upmu\text{s}$, comparable to the root-mean-square noise in J1603$-$7202's timing residuals \citep{Reardon+21}. However, the low inclination angle means the Shapiro delay is particularly difficult to decouple from the delay resulting from orbital variations. The effective amplitude in the timing residuals will therefore be much smaller and so we do not expect the Shapiro delay to be measurable with pulsar timing using current observing instruments.

\subsection{Structures in individual secondary spectra}\label{sec:interesting_spectra}
In this section we discuss some secondary spectra in our dataset that exhibit interesting features, for example, discrete ``blobs" of power, potential arclets, and other distributions of power inconsistent with the model prediction. With the model prediction for the arc curvature at each observation, we contextualize certain features and, using simulated spectra, speculate on the possible scattering geometries that may have caused them.

A selection of interesting spectra are shown alongside their normalized profiles in Figure \ref{fig:interesting_arcs}. Overlaid in red are parabolas with curvatures predicted using the static anisotropic model with the maximum-likelihood parameter values. In some cases, the curvatures predicted from the model seem to be inconsistent with those of the arcs. This is a consequence of our noise modification (see Section \ref{sec:static}), which absorbs this unmodelled scatter so that we only end up measuring the ``average" values of the model parameters.

The top-left and top-middle observations are only two days apart and show what appears to be a cell of power ``detaching" from the arc and moving to higher $f_\lambda$. If this interpretation is correct, this would physically correspond to a cell of scattering plasma moving away from the line-of-sight (see the discussion on multiple component images in \citet{Walker+04}). It is possible to transform a secondary spectrum into an on-sky scattered image; however, there is an ambiguity in which side of the velocity vector the signal originated from \citep{Cordes+06}. Common features between independent secondary spectra such as these can be used to circumvent this: since there should be a corresponding evolution in the scattered image, superimposing the images may allow one to ``triangulate" the true location of the signal. As we now possess a robust velocity model, the possibility is open for this kind of analysis to be performed. Imaging the physical structures responsible for the extreme scattering at AU scales may provide valuable clues to how such overdensities form. Because J1603$-$7202 is in the strong scintillation regime (see Section \ref{sec:curvatures}), we expect it to exhibit self-interference in the extended scattered image. Indeed, we see evidence for this in the secondary spectra in the form of broad arcs and inverted arclets. The observation at MJD 55580.73 (bottom right) features what appear to be distinguishable inverted arclets, suggesting particularly strong self-interference. To be able to interpret the transformed secondary spectra as scattered images, these self-interference effects must first be removed using a phase retrieval algorithm \citep{Walker+08,Baker+21,Sprenger+20}.

The observation at MJD $55145.9$ (bottom left) appears to exhibit a very strong phase gradient, and a double arc, with the curvature predicted by the model coinciding with the brighter, higher curvature arc. There are several possible explanations for the additional arc, such as scattering from a separate screen or, if belonging to the same screen as the other arc, a different anisotropy angle and/or ISM velocity. Unfortunately, this is only a transient feature and so cannot be modelled.

The observation at MJD $54310.58$ (top right) shows a region of power inconsistent with both the single-epoch model prediction (shown in the figure) and the three-epoch model prediction. However, it appears to be separate from the power at lower $f_\lambda$, suggesting it is a transient feature. Its physical interpretation is unclear as there are many potential combinations of the screen parameters that can account for its position.

The observation at MJD $55750.68$ (bottom middle) shows a thin arc-like feature to the left of the fitted arc. Since the feature does not appear to pass through the origin, it is unlikely to be another scintillation arc. Using the simulation routines included in \textsc{scintools}, based on code by \citet{Coles+10}, a similar feature is reproducible and is explained by an uptick in power at a particular point along the inverted arclets. An example simulated spectrum is shown in Figure \ref{fig:arc_sim}. The individual arclets themselves do not appear resolvable in this spectrum but maybe contribute to the broadening of the arc.

\section{Conclusion}
We modelled more than a decade of time variations in the arc curvature of binary PSR J1603$-$7202 by treating the scattering as being from a thin screen of plasma between the Earth and the pulsar. The data are well-fit by a single-screen model. However, we see moderately strong evidence for time-variability in the parameters of the scattering medium, with the preferred models possessing three or more epochs during which the screen is described by a different set of parameters. 

The inclination angle and longitude of ascending node of the pulsar's orbit are consistent with results from pulsar timing but provide significantly better constraints. This illustrates the power of scintillation modelling to supplement timing for low-inclination orbits. We also measured a fractional distance to the screen of $s=0.25\substack{+0.09 \\ -0.08}$, which does not appear to coincide with any catalogued star in the vicinity of the pulsar's on-sky position. This leaves open the question: what was the source of the extreme scattering event observed for J1603$-$7202?

From our measurement of the inclination angle, we place a lower bound on the mass of J1603$-$7202's companion of $\gtrsim 0.5\,\text{M}_\odot$ assuming a pulsar mass of $\gtrsim 1.2\,\text{M}_\odot$. This would place the companion in the rarer class of massive white dwarfs and likely rules out it being a helium white dwarf. This lower bound on the companion mass further gives a lower bound on the expected peak Shapiro delay of $\gtrsim 1.6\,\upmu\text{s}$, comparable to the RMS noise in J1603$-$7202's timing residuals. However, due to the low inclination angle this signal will be greatly reduced in the post-fit residuals and is unlikely to be measurable with current instruments.

For future work, it will be interesting to see how the results obtained here compare with those from modelling the variations in the decorrelation bandwidth and scintillation timescale. Modelling these alongside the curvature variations simultaneously may further improve the parameter measurements.

\section*{Acknowledgements}
The data used in this work were acquired as part of the PPTA project. We thank our PPTA colleagues for contributing to the observations and George Hobbs for making our data available on the CSIRO Data Access Portal (DAP). The authors are supported through Australian Research Council (ARC) Centre of Excellence CE170100004.  Parkes radio telescope (Murriyang) is part of the Australia Telescope, which is funded by the Commonwealth Government for operation as a National Facility managed by CSIRO. This research has made use of NASA's Astrophysics Data System and the ATNF Pulsar Catalogue.

%%%%%%%%% REFERENCES %%%%%%%%%

\bibliography{references.bib} 
\newpage
\appendix
\begin{figure}[H]
    \centering
    \includegraphics[width=1.0\textwidth]{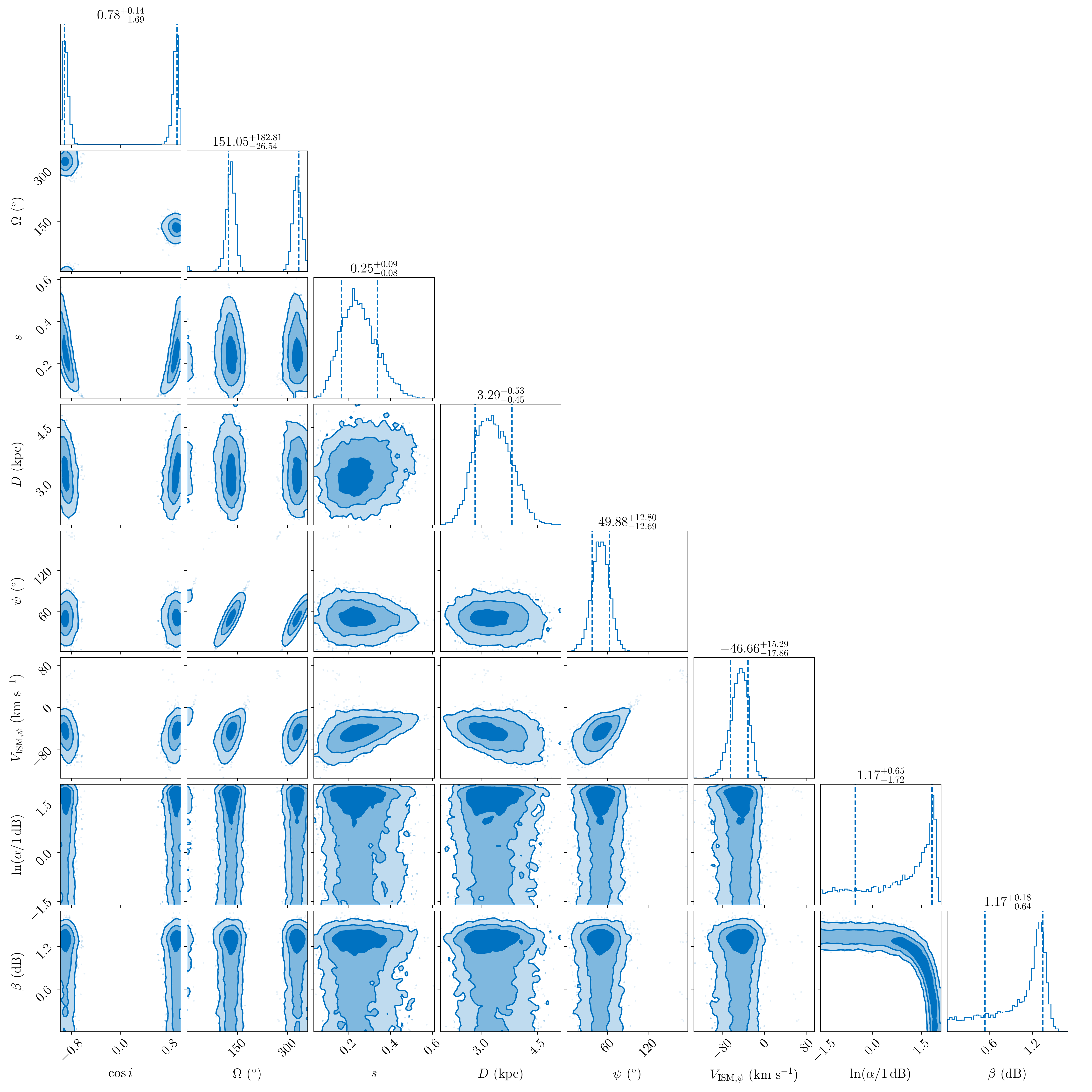}\\
    \caption{Corner plot for the anisotropic single-epoch model showing the marginalized 1D and 2D posterior distributions. The median and $1\sigma$ confidence intervals are given at the top of each column, however some may be unreliable due to degenerate solutions or significant non-Gaussianity. The complete list of parameter values is given in Table \ref{tab:params}.}
    \label{fig:corner}
\end{figure}

\begin{figure}[H]
    \centering
    \includegraphics[width=0.9\textwidth]{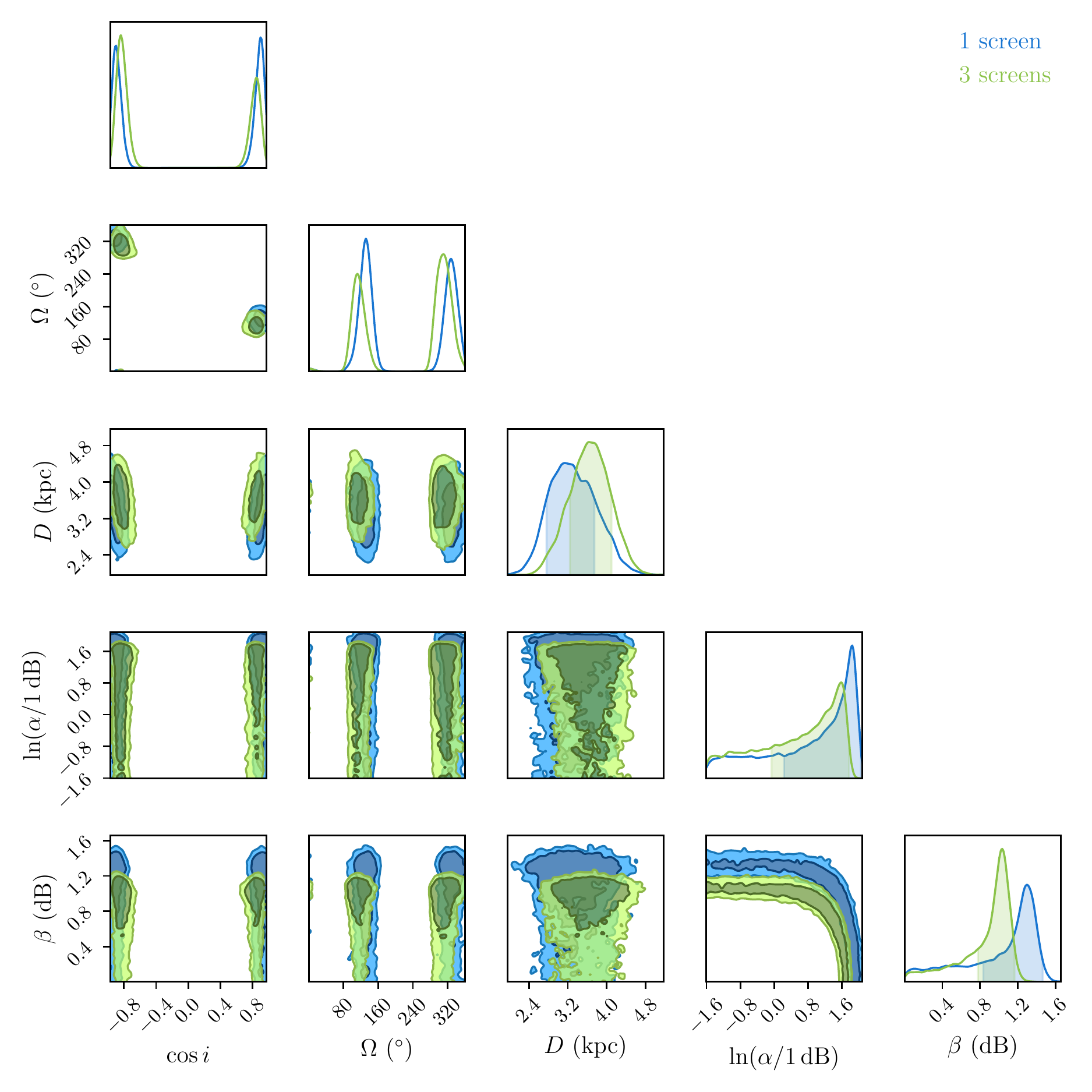}\\
    \caption{Comparison between the posteriors of the time-independent parameters for the anisotropic single-epoch and three-epoch model.}
    \label{fig:tvary_corner}
\end{figure}

\begin{figure}[H]
    \centering
    \includegraphics[width=0.8\textwidth]{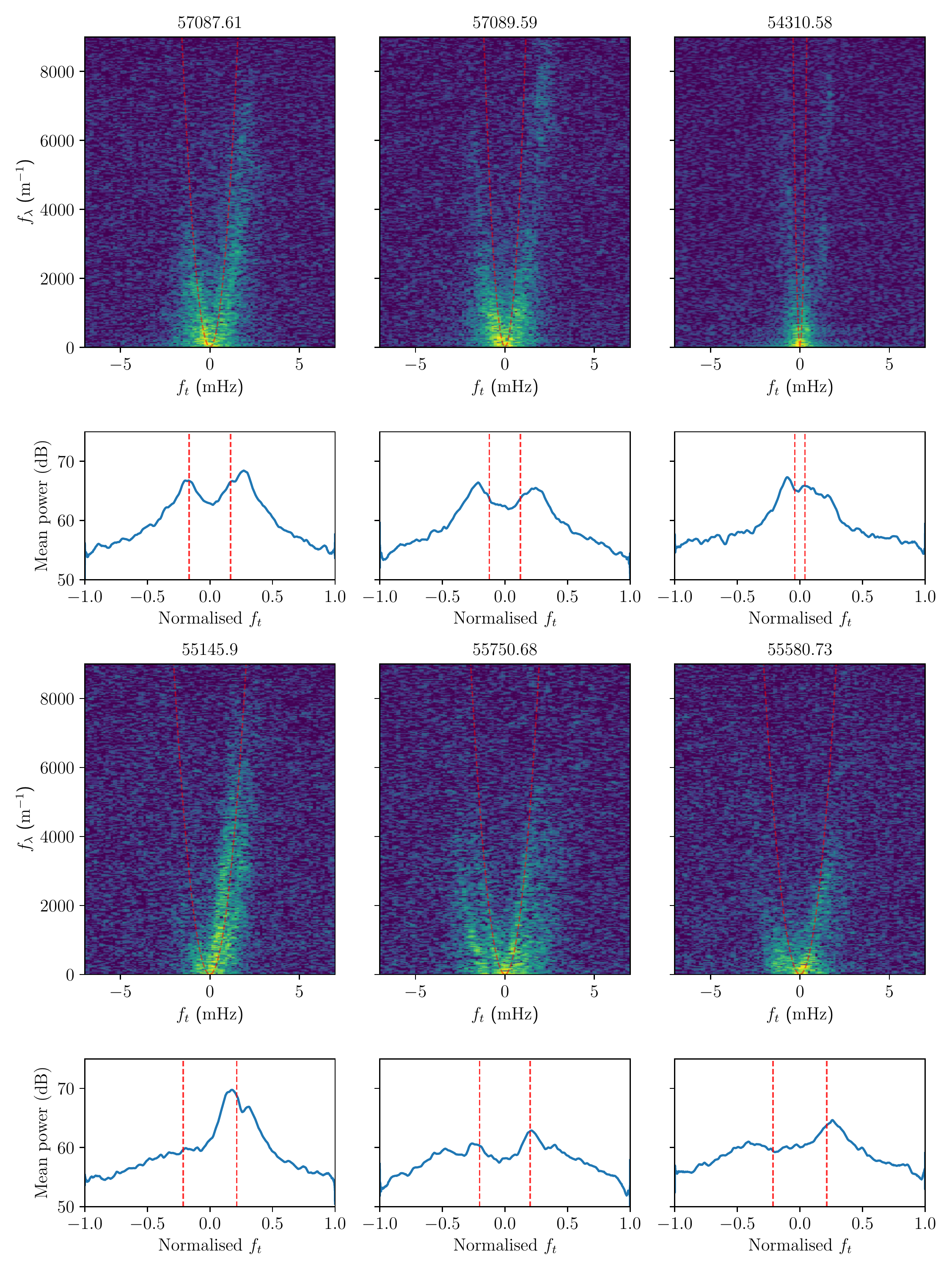}
    \caption{A selection of secondary spectra displaying interesting features (discussed in-text) alongside their corresponding normalized curvature profiles. The plots are titled by the MJD date of observation. The red, dashed parabolas have curvatures predicted by the anisotropic single-screen model using the maximum-likelihood parameters. In some cases, the curvatures predicted from the model appear to be inconsistent with those of the arcs, which is a consequence of our noise modification (discussed in Section \ref{sec:static}) absorbing this unmodelled scatter so that we only measure the ``average" values of the model parameters.}
    \label{fig:interesting_arcs}
\end{figure}

\begin{figure}[H]
    \centering
    \includegraphics[width=0.4\textwidth]{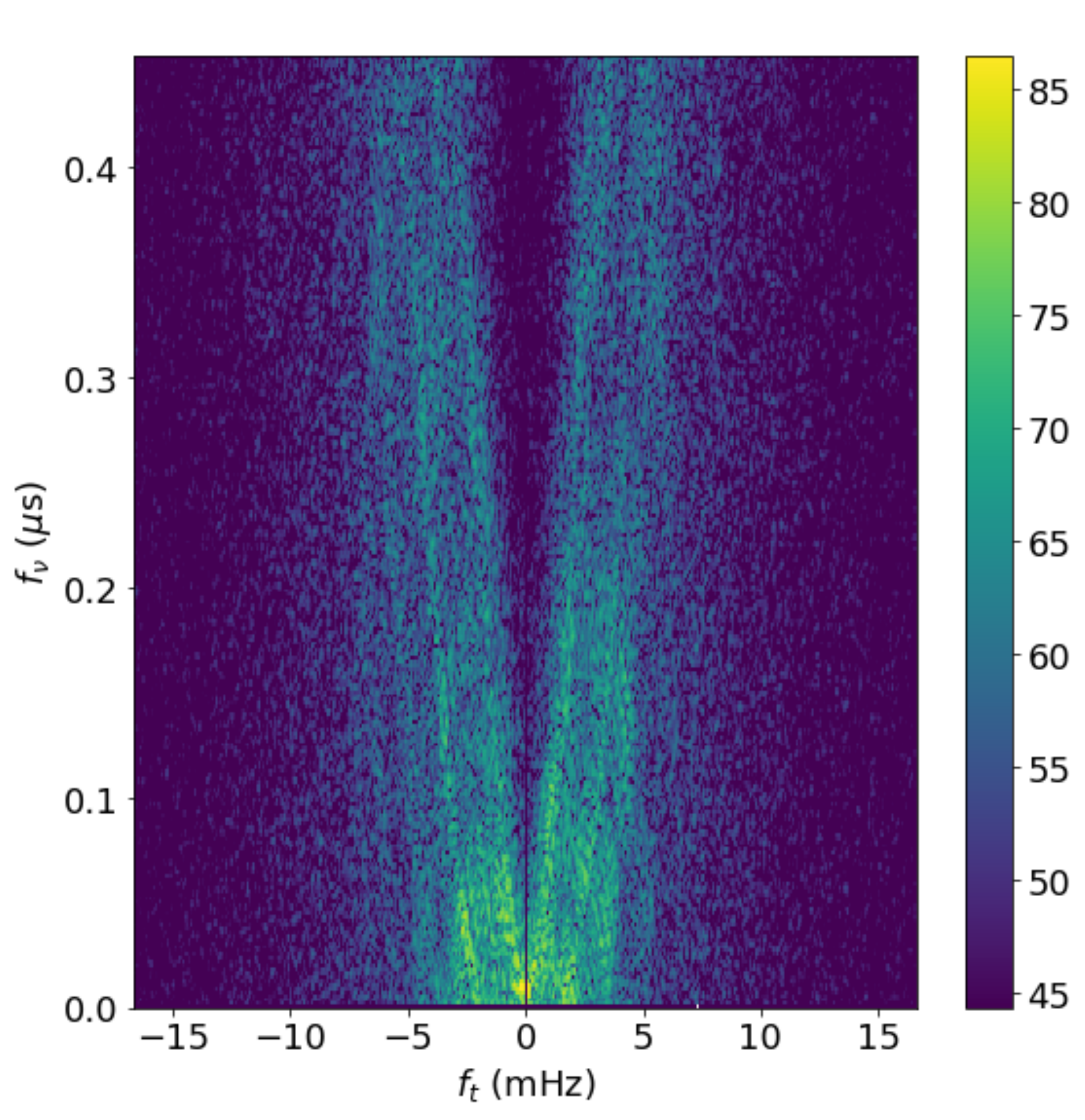}
    \caption{Simulated secondary spectrum using the technique described in \citet{Coles+10}, with $m_\text{B}^2=80$ and an anisotropy axial ratio of 4. A feature similar to that seen left of the arc in the bottom left panel of Figure \ref{fig:interesting_arcs} is present, attributed to an increase in power at a particular point along the inverted arclets.}
    \label{fig:arc_sim}
\end{figure}

\section{Accessing data and reproducing results}\label{sec:accessing_data}
The dynamic spectra for J1603$-$7202 are available for download from the CSIRO data access portal (DAP) at \url{https://doi.org/10.25919/82f5-mh79}. The process for generating these spectra from the raw observations is described in \citet{Kerr+20}. Code for reproducing and visualizing the results presented in this paper can be found at \url{https://github.com/kriswalker/J1603-7202_analysis}.

\end{document}